\documentclass[aps,prd,amsmath
,eqsecnum 
,preprintnumbers
,nofootinbib 
 ,showpacs 
 ,showkeys 
,twocolumn 
]{revtex4}
\usepackage[dvips]{graphicx}
\usepackage{amsmath}
\usepackage{amsfonts}
\usepackage{amssymb}
\usepackage{longtable} 
\usepackage{color}

\allowdisplaybreaks[1] 

\newcommand{\df}{\ {\overset {\rm def} =}\ }
\newcommand{\dr}[2]{\frac {{\rm d} {#1}} {{\rm d} {#2}}}

\newcommand{\dril}[2]{{{\rm d} {#1}} / {{\rm d} {#2}}}

\newcommand{\llim}[1] {\ {\underset {#1} {\longrightarrow}}\ }

\begin{document}

\title{Existence of blueshifts in quasi-spherical Szekeres spacetimes}

\author{Andrzej Krasi\'nski}
\affiliation{N. Copernicus Astronomical Centre, Polish Academy of Sciences, \\
Bartycka 18, 00 716 Warszawa, Poland} \email{akr@camk.edu.pl}

\date {}

\begin{abstract}
In Lema\^{\i}tre -- Tolman (L--T) models, light rays emitted \textit{radially}
at the Big Bang (BB) at such radial coordinates $r$ where the bang-time function
$t_B(r)$ has $\dril {t_B} r \neq 0$ reach every observer with \textit{infinite
blueshift}, $z = -1$. Consequently, there exist rays, emitted soon after the BB,
that will reach later observers with finite blueshift ($-1 < z < 0$). But in
spacetimes without symmetry there are no radial directions. The question thus
arises whether blueshifts can exist at all in the Szekeres models that contain
L--T as a limit, but in general have no symmetry. The aim of the present paper
is to show that strong blueshifts can be generated in quasi-spherical Szekeres
(QSS) models. It is shown that in an axially symmetric QSS model, infinite
blueshift can appear only on axial rays, which intersect every space orthogonal
to the dust flow on the symmetry axis. In an exemplary QSS model it is
numerically shown that if such a ray is emitted from the Big Bang where $\dril
{t_B} r \neq 0$, then indeed observers see it with $z \approx -1$. Rays emitted
shortly after the BB and running close to the symmetry axis will reach the
observers with a strong blueshift, too. Then, in a toy QSS model that has no
symmetry, it was shown by numerical calculations that two null lines exist such
that rays in their vicinity have redshift profiles similar to those in a
vicinity of the axial rays in the axially symmetric case. This indicates that
rays generating infinite blueshifts exist in general QSS spacetimes and are
concentrated around two directions.
\end{abstract}

\maketitle

\section{Motivation and background}\label{intro}

\setcounter{equation}{0}

In the spherically symmetric Lema\^{\i}tre \cite{Lema1933} -- Tolman
\cite{Tolm1934} (L--T) cosmological models some of the radial null geodesics
have a peculiar property. Suppose the point $P_o$ on the radial null geodesic G
lies later than the Big Bang (BB), and we follow G to the past until it
intersects the BB at $P_e$. The redshift $z$ observed at $P_o$ depends on the
slope of the BB at $P_e$. Namely, if the bang-time function $t_B(r)$ has $\dril
{t_B} r \neq 0$ at $P_e$, then $z(P_o) = -1$. On all other geodesics (G being
nonradial \textit{or} $\dril {t_B} r = 0$ at $P_e$), $z(P_o) = \infty$
\cite{Szek1980,HeLa1984,Kras2016}. The $z < 0$ property, referred to as
blueshift, means that the observed frequency $\nu_o$ of an electromagnetic wave
is greater than the emitted frequency $\nu_e$, and $z \to -1$ implies $\nu_o \to
\infty$. Since the L--T model is unrealistic before the last-scattering
hypersurface (LSH), in the real Universe $z(P_o)$ would always be greater than
$-1$, and $\nu_o$ would always be finite.

In L--T models the meaning of a radial or nonradial direction is obvious. Not so
in Szekeres models \cite{Szek1975,Szek1975b,PlKr2006}, which in general have no
symmetry. But the L--T models are contained in those of Szekeres as a
spherically symmetric limit. So, it is an interesting question whether $z < 0$
can arise also in Szekeres models, and on which rays. This is the subject of the
present paper. Only the quasi-spherical Szekeres (QSS) models are considered
here because the quasiplane and quasihyperbolic models are still poorly
understood \cite{HeKr2008,Kras2008,KrBo2012}. It is shown that the extra
flexibility provided by the dipole component of mass density in the Szekeres
models makes it easier to get strong blueshifts along certain directions while
the rays that generate those blueshifts are fewer. Numerical experiments showed
that such rays do exist in the axially symmetric case; they intersect each space
of constant $t$ on the symmetry axis.

The models considered here are not related to the actual cosmological
observations; they are meant to illustrate the improvements in generating
blueshifts achieved in comparison with the L--T models.

In Sec. \ref{QSSS}, properties of the QSS models are briefly described. In Sec.
\ref{origin}, the notion of an origin is defined and the behaviour of the
arbitrary functions of the QSS model at the origin is discussed. In Sec.
\ref{useform}, several useful formulae are listed for reference. In Sec.
\ref{nullgeo}, the equations of null geodesics in a general QSS spacetime are
introduced, and basic properties of those geodesics are briefly discussed. In
Sec. \ref{qsredshift}, the behaviour of redshift along null geodesics is
discussed, and necessary conditions for infinite blueshifts are derived.

In Sec. \ref{symmetric}, equations of null geodesics in an axially symmetric QSS
are displayed. In Sec. \ref{ERS}, the equation of an extremum redshift surface
is derived for those null geodesics that proceed along the symmetry axis in an
axially symmetric QSS (they will be called axial). In Sec. \ref{exQSS}, an
example of an axially symmetric QSS metric is introduced, and properties of
axial null geodesics in it are discussed in Sec. \ref{simpleblue}. These
geodesics do display strong blueshifts (SB) (i.e. $z \approx -1$ on them) if
they originate at the BB at points where $\dril {t_B} r \neq 0$. In Sec.
\ref{moreblue}, such null geodesics in the same QSS metric are discussed that
run close to the symmetry axis. They can generate strong blueshifts when they
originate close to the BB. (When they originate exactly at the BB, all observers
see infinite redshifts, independently of $\dril {t_B} r$.)

In Sec. \ref{simgen}, an exemplary QSS model is introduced that has no symmetry.
By numerical calculations it is shown that such rays exist in this model, on
which the redshift profiles have the same shape as on the rays running close to
the symmetry axis in the axially symmetric model. In particular, light sources
lying on those rays close to, but later than the BB, would be observed with
finite blueshifts. This is an indication that rays generating SB from the BB
should exist in their vicinity. However, in all known cases rays with SB are
unstable: an arbitrarily small perturbation of the observation point or
direction changes SB into infinite redshift. Therefore, for tracking such rays
numerically we must know how to keep them exactly on their unstable paths (as is
the case with radial rays in the L--T models and with axial rays in the axially
symmetric QSS models).

Section \ref{sumup} contains a summary of the results. Details of some
calculations are explained in the appendices.

\section{The quasispherical Szekeres spacetimes}\label{QSSS}

\setcounter{equation}{0}

The metric of the quasispherical Szekeres spacetimes \cite{Hell1996} can be
written as (notation adapted to \cite{PlKr2006})
\begin{equation}\label{2.1}
{\rm d} s^2 = {\rm d} t^2 - \frac {\left(\Phi,_r - \Phi {\cal E},_r/{\cal
E}\right)^2} {1 + 2 E(r)} {\rm d} r^2 - \left(\frac {\Phi} {\cal E}\right)^2
\left({\rm d} x^2 + {\rm d} y^2\right),\ \ \ \ \
\end{equation}
where
\begin{equation}\label{2.2}
{\cal E} \df \frac S 2 \left[\left(\frac {x - P} S\right)^2 + \left(\frac {y -
Q} S\right)^2 + 1\right],
\end{equation}
$P(r)$, $Q(r)$, $S(r)$ and $E(r)$ being arbitrary functions such that $S \neq 0$
and $E \geq -1/2$ at all $r$ ($E = -1/2$ can occur at isolated values of $r$,
but not on open intervals \cite{HeKr2002}).

The source in the Einstein equations is dust ($p = 0$), and the coordinates of
(\ref{2.1}) are comoving, so the velocity field of the dust is $u^{\alpha} =
{\delta_0}^{\alpha}$. The surfaces of constant $t$ and $r$ are nonconcentric
spheres, and $(x, y)$ are the stereographic coordinates on each sphere. At a
fixed $r$, the relation between $(x, y)$ and the spherical coordinates is
\begin{eqnarray}\label{2.3}
x &=& P + S \tan(\vartheta/2) \cos \varphi, \nonumber \\
y &=& Q + S \tan(\vartheta/2) \sin \varphi.
\end{eqnarray}
The functions $(P, Q, S)$ determine the positions of the centres of the spheres
in the spaces of constant $t$ (see an example in Sec. \ref{simgen}). The
function $\Phi(t,r)$ is determined by the same evolution equation as in the L--T
models:
\begin{equation}\label{2.4}
{\Phi,_t}^2 = 2 E(r) + \frac {2 M(r)} {\Phi} - \frac 1 3 \Lambda \Phi^2,
\end{equation}
where $\Lambda$ is the cosmological constant and $M(r)$ is an arbitrary
function. Any solution of (\ref{2.4}) depends on $t$ through the combination $(t
- t_B(r))$, where $t_B(r)$ is still one more arbitrary function; $t = t_B(r)$ is
the BB time, at which $\Phi(t_B, r) = 0$.

In the following, we shall assume $\Phi,_t > 0$ (the Universe is expanding) and
$\Lambda = 0$. The solutions of (\ref{2.4}) under these assumptions are
presented in Appendix \ref{freqs}; they are the same as the Friedmann solutions
\cite{PlKr2006}.

The mass density implied by (\ref{2.1}) is
\begin{equation}\label{2.5}
\kappa \rho = \frac {2 \left(M,_r - 3 M {\cal E},_r / {\cal E}\right)} {\Phi^2
\left(\Phi,_r - \Phi {\cal E},_r / {\cal E}\right)}, \quad \kappa \df \frac {8
\pi G} {c^2}.
\end{equation}

In choosing the arbitrary functions, one must take care that the resulting mass
density in the region being considered is positive and finite. The conditions
that ensure this were worked out in Ref. \cite{HeKr2002}, and they are:
\begin{eqnarray}
\frac {M,_r} {3M} &\geq& \frac {\sqrt{(S,_r)^2 + (P,_r)^2 + (Q,_r)^2}} S
~~~~\forall~r, \label{2.6} \\
\frac {E,_r} {2E} &>& \frac {\sqrt{(S,_r)^2 + (P,_r)^2 + (Q,_r)^2}} S
~~~~\forall~r. \label{2.7}
\end{eqnarray}
These inequalities ensure that \cite{HeKr2002}
\begin{equation}\label{2.8}
\frac {M,_r} {3M} \geq \frac {{\cal E},_r} {\cal E}, \qquad \frac {E,_r} {2E} >
\frac {{\cal E},_r} {\cal E} \qquad \forall~r.
\end{equation}

As first noted by Szekeres \cite{Szek1975b} and elaborated by de Souza
\cite{DeSo1985}, the density distribution (\ref{2.5}) is that of a mass-dipole
superposed on a spherically symmetric monopole. The dipole contribution is
generated by the term ${\cal E},_r/{\cal E}$ and vanishes on the set where
${\cal E},_r = 0$. The extrema of density coincide with the extrema of ${\cal
E},_r/{\cal E}$: the density is minimum where ${\cal E},_r/{\cal E}$ is maximum
and vice versa \cite{HeKr2002}.

\section{The origin and the behaviour of the arbitrary functions at
it}\label{origin}

\setcounter{equation}{0}

It is not necessary for a Szekeres spacetime to have an origin. Spacetimes
without an origin have cylindrical topology of the constant-time subspaces; in
the spherically symmetric limit they do not contain the center of symmetry. This
configuration is somewhat exotic, so we shall assume that an origin exists. It
is the set at which each sphere of constant $t$ and $r$ in (\ref{2.1}) has zero
radius, i.e. where $\Phi = 0$ at all $t > t_B$. Multiplying (\ref{2.4}) by
$\Phi$, anticipating that $\left|\left.\Phi,_t\right|_{\rm origin}\right| <
\infty$, and then taking the result at the origin we obtain that
\begin{equation}\label{3.1}
M_{\rm origin} = 0.
\end{equation}
Since $M$ depends only on $r$, the origin worldline is a line of constant $r$,
i.e. the origin is comoving and coincides always with the same dust particle.

The metric (\ref{2.1}) is covariant with the transformations $r = f(r')$, where
$f(r')$ is an arbitrary function. These can be used to give one of the arbitrary
functions a convenient shape. It is advantageous to choose $r$ so that
\begin{equation}\label{3.2}
M(r) = M_0 r^3,
\end{equation}
where $M_0$ is a constant, which can be given an arbitrary nonzero value by a
further transformation $r' = A r''$, where $A$ is a constant. We shall assume
(\ref{3.2}) and $M_0 = 1$, but $M_0$ will be kept in all formulae to avoid
confusion about dimensions of various derived quantities.

Equations (\ref{3.1}) and (\ref{3.2}) imply
\begin{equation}\label{3.3}
r_{\rm origin} = 0.
\end{equation}

We assume\footnote{At points where $S = 0$ we have $\Phi/{\cal E} = 0$ in
(\ref{2.1}), so they are additional origins. The Riemann tensor will be finite
at those points if $|{\cal E},_r| < \infty$ there, see Appendix \ref{riemann}.
The metric does not change under the substitution $S = - {\cal S}$, so the
assumption $S > 0$ is not a limitation.}
\begin{equation}\label{3.4}
S > 0 \ {\rm everywhere} \Longrightarrow {\cal E} > 0 \ {\rm everywhere}.
\end{equation}

Since at $r = 0$ both $M = 0$ and $\Phi = 0$, using (\ref{3.4}) one can
calculate the limit

\parbox{8.3cm}{\vspace{-3mm}
\begin{eqnarray}\label{3.5}
&& \lim_{r \to 0} \frac {\Phi^3} M \equiv \lim_{r \to 0} \frac {(\Phi / {\cal
E})^3} {M / {\cal E}^3} = \lim_{r \to 0} \frac {3 \Phi^2 \left(\Phi,_r - \Phi
{\cal E},_r / {\cal E}\right)} {M,_r - 3 M {\cal E},_r / {\cal E}} \nonumber \\
&& \equiv \left.\frac 6 {\kappa \rho(t, r)}\right|_{r = 0} \df \frac {R^3(t)}
{M_0}.
\end{eqnarray}
 }
This limit is thus finite at those points of the origin where $\rho(t, 0) \neq
0$. At these points, $\Phi$ must have the form
\begin{equation}\label{3.6}
\Phi = r [R(t) + {\cal B}(t,r)],
\end{equation}
where ${\cal B}$ has the property
\begin{equation}\label{3.7}
\lim_{r \to 0}{\cal B} = 0.
\end{equation}
Note that no approximation is involved in (\ref{3.6}); this is a
reparametrisation of $\Phi$ that respects (\ref{3.5}). In the Friedmann limit
${\cal B} \equiv 0$, and $R$ becomes the scale factor.

Using (\ref{3.6}) and (\ref{3.2}) in (\ref{2.4}) we conclude that $E$ must have
the form
\begin{equation}\label{3.8}
2 E = r^2 (- k + {\cal F}(r)),
\end{equation}
where $k$ is a constant and ${\cal F}(r)$ has the property
\begin{equation}\label{3.9}
\lim_{r \to 0}{\cal F} = 0.
\end{equation}
Equation (\ref{3.8}) is also an exact formula. In the Friedmann limit ${\cal F}
\equiv 0$, and $k$ becomes the curvature index.

\section{Some useful formulae}\label{useform}

\setcounter{equation}{0}

In the next sections we shall need to know the behaviour of $\Phi,_r$ and
$\Phi,_{tr}$ at the BB and at $r \to 0$. We have (\cite{Kras2016}, Eqs. (3.9)
and (3.10)):

\begin{eqnarray}
&& \Phi,_r = \left(\frac {M,_r} M - \frac {E,_r} E\right)\Phi
\nonumber \\
&& + \left[\left(\frac {3 E,_r} {2 E} - \frac {M,_r} M\right) \left(t -
t_B\right) - t_{B,r}\right] \Phi,_t, \ \ \ \ \ \label{4.1} \\
&& \Phi,_{tr} = \frac {E,_r} {2E}\ \Phi,_t \nonumber \\
&& - \frac M {\Phi^2}\ \left[\left(\frac {3 E,_r} {2 E} - \frac {M,_r} M\right)
\left(t - t_B\right) - t_{B,r}\right]. \ \ \ \ \ \label{4.2}
\end{eqnarray}

\subsection{Limits at the BB}

These limits are calculated at $r > 0$. Limits at the points where
simultaneously $t = t_B$ and $r = 0$ are nonunique: they depend on the detailed
shapes of the arbitrary functions and on the path of approach to such a point;
see an example for the L--T model in Ref. \cite{PlKr2006}.

With $t \to t_B$ (so $\Phi \to 0$) we find using (\ref{2.4}):
\begin{eqnarray}
&& \lim_{t \to t_B} \Phi,_t = \lim_{t \to t_B} \Phi,_{tr} = \infty, \label{4.3}
\\
&& \lim_{t \to t_B} \left[\left(t - t_B\right) / \Phi\right] = 0, \label{4.4} \\
&& \lim_{t \to t_B} \left[(t - t_B) \Phi,_t\right] = \lim_{t \to t_B} \left(\Phi
\Phi,_t\right) = 0, \label{4.5} \\
&& \lim_{t \to t_B} \left[\left(t - t_B\right) \Phi,_t/\Phi\right] = \frac 2 3.
\label{4.6}
\end{eqnarray}
{}From (\ref{4.3}) and (\ref{4.5}) we find
\begin{equation}\label{4.7}
\lim_{t \to t_B} \Phi,_r = - t_{B,r} \lim_{t \to t_B} \Phi,_t.
\end{equation}
Thus, on a curve that hits the BB where $\dril {t_B} r \neq 0$ we have $\lim_{t
\to t_B} \Phi,_r = \pm \infty$; the sign in front of $\infty$ is the sign of $(-
\dril {t_B} r)$. On a curve that hits the BB where $\dril {t_B} r = 0$ we have
$\lim_{t \to t_B} \Phi,_r = 0$.

Further, it follows from (\ref{4.6}) that
\begin{equation}\label{4.8}
\lim_{t \to t_B} \left(\frac {\Phi,_r} {\Phi}\right) = \frac {M,_r} {3 M} -
t_{B,r} \lim_{t \to t_B}\left(\frac {\Phi,_t} {\Phi}\right).
\end{equation}
This limit is finite when $\dril {t_B} r = 0$ at the intersection of the path of
approach with the BB, and is infinite otherwise; the sign of the infinity is
again the sign of $\left(- t_{B,r}\right)$.

\subsection{Limits at the origin}

The limits given below are calculated at $t > t_B$; the reason for avoiding the
point $(r, t) = (0, t_B)$ is the same as in the previous subsection.

Using (\ref{2.4}), (\ref{3.2}) and (\ref{3.6}) -- (\ref{3.8}) we find:
\begin{eqnarray}
&& \lim_{r \to 0} \Phi,_t = 0, \label{4.9} \\
&& \lim_{r \to 0} \frac {\Phi,_t} {\Phi} = \frac {R,_t} R < \infty, \label{4.10}
\\
&& \lim_{r \to 0} \Phi,_r = R < \infty, \label{4.11} \\
&& \lim_{r \to 0} \Phi,_r / \Phi = \infty, \label{4.12}
\end{eqnarray}
\begin{eqnarray}
&& \lim_{r \to 0} \Phi,_{tr} = R,_t < \infty. \label{4.13}
\end{eqnarray}

\section{Null geodesics in quasispherical Szekeres spacetimes}\label{nullgeo}

\setcounter{equation}{0}

We denote
\begin{equation}\label{5.1}
\left(k^t, k^r, k^x, k^y\right) \df \dr {(t, r, x, y)} {\lambda},
\end{equation}
where $\lambda$ is the affine parameter, and
\begin{equation}\label{5.2}
{\cal N} \df \Phi,_r - \Phi {\cal E},_r/{\cal E}.
\end{equation}
Then the equations of geodesics for (\ref{2.1}) are \cite{BKHC2010}
\begin{eqnarray}
\dr {k^t} {\lambda} &+& \frac {{\cal N} {\cal N},_t} {1 + 2E} \left(k^r
\right)^2 + \frac{\Phi {\Phi,_t}}{{\cal E}^2} \left[\left(k^x\right)^2 +
\left(k^y \right)^2\right] = 0, \ \ \ \ \ \ \ \label{5.3} \\
\dr {k^r} {\lambda} &+& 2 \frac {{\cal N},_t} {\cal N} k^t k^r +
\left(\frac{{\cal N},_r} {{\cal N}} - \frac {E,_r} {1 + 2E}\right)
\left(k^r\right)^2 \nonumber \\
&-& 2 \Phi \ \frac {({\cal E},_r / {\cal E}),_x k^x + ({\cal E},_r / {\cal
E}),_y k^y} {\cal N} k^r \nonumber \\
&-& \frac {\Phi} {{\cal E}^2} \frac {1 + 2E} {{\cal N}} \left[\left(k^x\right)^2
+ \left(k^y\right)^2\right] = 0, \label{5.4} \\
\dr {k^x} {\lambda} &+& 2 \frac {\Phi,_t} {\Phi} k^t k^x + \frac {{\cal E}^2
{\cal N}} {\Phi (1 + 2E)}\ \left(\frac {{\cal E},_r} {\cal E}\right),_x
\left(k^r\right)^2 \nonumber \\
&+& 2 \frac {\cal N} {\Phi} k^r k^x - \frac { {\cal E},_x}{{\cal E}}
\left(k^x\right)^2 \nonumber \\
&-& 2 \frac {{\cal E},_y} {\cal E} k^x k^y + \frac{{\cal E},_x} {\cal E}
\left(k^y\right)^2 = 0, \label{5.5} \\
\dr {k^y} {\lambda} &+& 2 \frac {\Phi,_t} {\Phi} k^t k^y + \frac {{\cal E}^2
{\cal N}} {\Phi (1 + 2E)}\ \left(\frac {{\cal E},_r} {\cal E}\right),_y
\left(k^r\right)^2 \nonumber \\
&+& 2 \frac {\cal N} {\Phi} k^r k^y + \frac {{\cal E},_y} {\cal E}
\left(k^x\right)^2 \nonumber \\
&-& 2 \frac {{\cal E},_x} {\cal E} k^x k^y - \frac {{\cal E},_y} {\cal E}
\left(k^y\right)^2 = 0. \label{5.6}
\end{eqnarray}
The geodesics determined by (\ref{5.3}) -- (\ref{5.6}) are null when
\begin{equation}\label{5.7}
\left(k^t\right)^2 - \frac {{\cal N}^2 \left(k^r\right)^2} {1 + 2E(r)} -
\left(\frac {\Phi} {\cal E}\right)^2 \left[\left(k^x\right)^2 +
\left(k^y\right)^2\right] = 0.
\end{equation}

Note that $k^r \neq 0$ over any open interval of a null geodesic: otherwise,
$\dril {k^r} {\lambda} = 0$ in that interval, and (\ref{5.4}) would imply $k^x =
k^y = 0$; such a geodesic would be timelike. However, $k^r = 0$ is allowed at
isolated points. Thus, $r$ can be used as a parameter on any arc of a null
geodesic, on which $k^r$ does not change sign.

Let the subscript $o$ refer to the observation point. We will mostly consider
past-directed rays, on which $k^t < 0$. Since the affine parameter along each
single geodesic is determined up to the transformations $\lambda = a \lambda' +
b$, where $a$ and $b$ are constants, it can be chosen such that
\begin{equation}\label{5.8}
k^t_o = -1,
\end{equation}
and this choice will be made throughout this paper. Then, from (\ref{5.7}) we
have
\begin{equation}\label{5.9}
\left(k_o^x\right)^2 + \left(k_o^y\right)^2 \leq \left(\frac {{\cal E}_o}
{\Phi_o}\right)^2;
\end{equation}
the equality occurs only when $k_o^r = 0$, i.e. when the null geodesic passes
through the observation event tangentially to the hypersurface of constant $r$.

The coefficient ${\cal N}/\Phi$ in front of $(k^r)^2$ and of $k^r$ in
(\ref{5.5}) and (\ref{5.6}) becomes infinite at the origin, as follows from
(\ref{4.12}). Thus, when running a numerical calculation of a geodesic through
the origin the limits of $\left({\cal E},_r / {\cal E}\right),_x
\left(k^r\right)^2 / \Phi$, $\left({\cal E},_r / {\cal E}\right),_y
\left(k^r\right)^2 / \Phi$, $k^r k^x / \Phi$ and $k^r k^y / \Phi$ have to be
evaluated exactly. If they are finite, then it is best to choose the origin as
the initial point. See Sec. \ref{moreblue} for an example.

\section{Redshift in the quasispherical Szekeres spacetimes}\label{qsredshift}

\setcounter{equation}{0}

\setcounter{table}{0}

The general formula for redshift along a ray emitted at $P_e$ and observed at
$P_o$ is \cite{Elli1971}
\begin{equation}\label{6.1}
1 + z = \frac {\left(u_{\alpha} k^{\alpha}\right)_e} {\left(u_{\alpha}
k^{\alpha}\right)_o},
\end{equation}
where $u_{\alpha}$ are the four-velocity vectors of the emitter and of the
observer, and $k^{\alpha}$ is the affinely parametrised tangent vector field to
the ray. In our case, both the emitter and the observer will be assumed to
comove with the cosmic matter, so $u_{\alpha} = {\delta^0}_{\alpha}$, and then
(\ref{6.1}) simplifies to $1 + z = {k_e}^t/{k_o}^t$. A further simplification
results when the affine parameter is rescaled so that (\ref{5.8}) holds; then
\begin{equation}\label{6.2}
1 + z = - {k_e}^t.
\end{equation}
Since $k^t = - \dril t {\lambda}$ along a past-directed ray, using (\ref{6.2})
we get from (\ref{5.7})
\begin{eqnarray}
\left(1 + z\right)^2 = \left\{\frac {{\cal N}^2 \left(k^r\right)^2} {1 + 2E(r)}
+ \left(\frac {\Phi} {\cal E}\right)^2 \left[\left(k^x\right)^2 +
\left(k^y\right)^2\right]\right\}_e. \nonumber \\ \label{6.3}
\end{eqnarray}

Denote
\begin{equation}\label{6.4}
\left(k^x\right)^2 + \left(k^y\right)^2 \df J^2.
\end{equation}
Then we obtain from (\ref{5.5}) -- (\ref{5.6})
\begin{eqnarray}\label{6.5}
&& \frac 1 J \dr J s + 2 \frac {\Phi,_t} {\Phi}\ k^t + 2 \frac {\Phi,_r} {\Phi}\
k^r \nonumber \\
&-& \frac {{\cal E},_r} {\cal E}\ k^r - \frac {{\cal E},_x} {\cal E}\ k^x -
\frac {{\cal E},_y} {\cal E}\ k^y = L(s),
\end{eqnarray}
where
\begin{eqnarray}
&& L(s) = L(t(s), r(s), x(s), y(s)) \nonumber \\
&& \df - \frac {\cal N} {\Phi (1 + 2E)}\ \frac {\left(k^r\right)^2 {\cal E}^2}
{J^2} {\cal D}_k + \frac {{\cal E},_r} {\cal E}\ k^r, \label{6.6} \\
&& {\cal D}_k \df \left[\left(\frac {{\cal E},_r} {\cal E}\right),_x k^x +
\left(\frac {{\cal E},_r} {\cal E}\right),_y k^y\right]. \label{6.7}
\end{eqnarray}
As can be verified using (\ref{2.2}), $({\cal E},_r / {\cal E}),_x$ and $({\cal
E},_r / {\cal E}),_y$ are finite for all values of $x$, $y$ and $r$.
Consequently, ${\cal D}_k$ is finite at all points where $k^x$ and $k^y$ are
finite.

Since $\Phi$ depends only on $t$ and $r$, while ${\cal E},_t = 0$, we have
$\Phi,_t k^t + \Phi,_r k^r = \dril {\Phi} s$ and ${\cal E},_r k^r + {\cal E},_x
k^x + {\cal E},_y k^y = \dril {\cal E} s$. Assuming that the initial condition
for (\ref{6.5}) is given at the observation point, where $s = s_o$, the solution
of (\ref{6.5}) may be written as
\begin{equation}\label{6.8}
J = \frac {\cal E} {{\cal E}_o} \left(\frac {\Phi_o} {\Phi}\right)^2\ J_o\
\exp\left(\int_{s_o}^s L(\lambda) {\rm d} \lambda\right).
\end{equation}
Substituting (\ref{6.8}) in (\ref{6.3}) we obtain
\begin{equation}\label{6.9}
(1 + z)^2 = \left[\frac {{\cal N}^2 \left(k^r\right)^2} {1 + 2E} + \frac
{{J_o}^2 {{\Phi_o}^4}} {{{\cal E}_o}^2 \Phi^2}\ \exp\left(2 \int_{s_o}^s
L(\lambda) {\rm d} \lambda\right)\right]_e.
\end{equation}
The geodesic would be radial if $J = J_o \equiv 0$, but such geodesics, as
mentioned above, do not exist in general.

Now suppose we follow the ray from the observation point back in time to its
intersection with the BB, where $\Phi \to 0$. Can (\ref{6.9}) allow for infinite
blueshift, i.e. for
\begin{equation}\label{6.10}
\lim_{t \to t_B} z \df z_{\rm BB} = -1?
\end{equation}

Both terms on the right-hand side of (\ref{6.9}) are non-negative, so to allow
$z_{\rm BB} = -1$ they both must go to zero when $t \to t_B$. As follows from
(\ref{4.7}), the first term will go to zero when $\dril {t_B} r = 0$ at the
intersection of the ray with the BB. But when $\dril {t_B} r \neq 0$ at that
point, then a necessary condition for $z_{\rm BB} = -1$ is
\begin{equation}\label{6.11}
\lim_{t \to t_B} k^r = 0.
\end{equation}

Nothing general can be said about the behaviour of the second term on the
right-hand side of (\ref{6.9}) when $\Phi_o = 0$, i. e. when the observation
point is at the origin. In the L--T limit, a ray passing through the origin is
radial, and the second term drops out. Here, however, $L$ at the origin is
infinite (because of $\Phi,_r/\Phi$, see (\ref{4.12})), so $\int_{s_o}^s L {\rm
d} \lambda$ can be infinite, too, and the exponential factor in (\ref{6.9}) can
compensate for $\Phi_o \to 0$. This can be investigated only numerically case by case.

If $\Phi_o \neq 0$, then, irrespectively of the behaviour of the first term, $1
+ z_{\rm BB} \to \infty$ as long as the coefficient of $\Phi^{-2}$ in the last
term has a nonzero limit at the BB. So, another necessary (but not sufficient)
condition for $z_{\rm BB} = -1$ is
\begin{equation}\label{6.12}
\frac {J_o {{\Phi_o}^2}} {{\cal E}_o} \lim_{t \to t_B} \left[\frac 1 {\Phi}\
\exp\left(\int_{s_o}^s L(\lambda) {\rm d} \lambda\right)\right] = 0.
\end{equation}
{}From (\ref{3.4}), ${\cal E} \neq 0$ everywhere, and is finite except at $x \to
\infty$ and $y \to \infty$ (which are coordinate singularities). So,
(\ref{6.12}) can be fulfilled (1) when $J_o = 0$, or (2) when
$\exp\left(\int_{s_o}^s L(\lambda) {\rm d} \lambda\right) \llim{t \to t_B} 0$
faster than $\Phi$.

Case (1) means that the null geodesic is orthogonal to the sphere of constant
$t$ and $r$ at the {\it observation} point. However, fulfilling (\ref{6.12}) in
this way is problematic: $L(\lambda)$ may be infinite at $\lambda = s_o$, and
this may cause that $\int_{s_o}^s L(\lambda) {\rm d} \lambda = \infty$ for any
$s_o$. There are too many possibilities to identify a criterion for $z_{\rm BB}
= -1$ in this case. Moreover, this way of achieving $z_{\rm BB} = -1$ would be
unnatural: the second term in (\ref{6.9}) would then vanish all along the ray
between the observation point and the BB. The implication would be that $z_{\rm
BB} = - 1$ if the ray is orthogonal to any single constant-$(t,r)$ surface,
independently of what happens between this surface and the BB.

A necessary condition for case (2) is
\begin{equation}\label{6.13}
\lim_{t \to t_B} L(s) = + \infty
\end{equation}
(because in integrating to the past $s < s_o$). Since ${\cal N} >0$ (in
consequence of the no-shell-crossing conditions \cite{HeKr2002}), and $1 + 2E
\geq 0$ (to have the right signature), the sign of the infinity in $L$ will be
determined by the sign of ${\cal D}_k$, which can be any.

One of the ways of fulfilling (\ref{6.13}) is $J \to 0$ at the BB (provided that
$k^r$ and ${\cal D}_k$ do not go to zero too fast). But whether the ray from the
BB is blue- or redshifted will depend here not only on the behaviour of $J$,
$k^r$ and $\dril {t_B} r$ near the BB, but also on whether the mass-dipole
component determined by ${\cal E},_r / {\cal E}$ is increasing or decreasing as
the (past-directed!) ray approaches the BB. When it decreases, ${\cal D}_k < 0$
and $\lim_{s \to s_{\rm BB}} L(s) = + \infty$, in which case (because of $s <
s_o$) $\exp \left(\int_{s_o}^s L(\lambda) {\rm d} \lambda\right) \to 0$, and
$z_{\rm BB} = -1$ is possible. But when it increases, ${\cal D}_k > 0$ and $\exp
\left(\int_{s_o}^s L(\lambda) {\rm d} \lambda\right) \to \infty$, preventing
$z_{\rm BB} = -1$.

Thus, unlike in the L--T case, formulating a clearcut criterion for infinite
bleshift in the Szekeres models is not possible, and the remarks above can only
be used as suggestions for numerical experiments.

\begin{table}[h]
\caption{Is $z = - 1$ possible at the BB? \label{treetable}}
\begin{tabular}{|c|c|}
 \hline \hline
$\left.\dr {t_B} r\right|_{\rm BB} = 0$ & \begin{tabular}{c c}
    $k^r_{\rm BB} = 0$ & \begin{tabular}{|c|c|l}
        $J_{\rm BB} = 0$ & P & I\ \ \ \ \ \\
        $J_{\rm BB} \neq 0$ & $z_{\rm BB} = \infty$ & II\ \ \ \ \\
        \end{tabular} \\
    \hline
    $k^r_{\rm BB} \neq 0$ & \begin{tabular}{|c|c|l}
        $J_{\rm BB} = 0$ & P & III\ \ \\
        $J_{\rm BB} \neq 0$ & $z_{\rm BB} = \infty$ & IV\ \ \ \\
        \end{tabular} \\
    \end{tabular} \\
 \hline
$\left.\dr {t_B} r\right|_{\rm BB} \neq 0$ & \begin{tabular}{c c}
    $k^r_{\rm BB} = 0$ & \begin{tabular}{|c|c|l}
        $J_{\rm BB} = 0$ & \ \ \ \ \ P \ \ \ \ \ & V\ \ \ \ \\
        $J_{\rm BB} \neq 0$ & \ \ \ \ \ P \ \ \ \ \ & VI\ \ \ \\
        \end{tabular} \\
    \hline
    $k^r_{\rm BB} \neq 0$ & \begin{tabular}{|c|c|l}
        $J_{\rm BB} = 0$ & $z_{\rm BB} = \infty$ & VII\ \hspace{-1.5mm} \\
        $J_{\rm BB} \neq 0$ & $z_{\rm BB} = \infty$ & VIII \hspace{-1.5mm} \\
        \end{tabular} \\
    \end{tabular} \\
 \hline \hline
\end{tabular}
\end{table}

By separately considering the various possibilities we arrive at Table
\ref{treetable}. The last column contains reference numbers of the cases. ``P''
stands for ``possibly $z_{\rm BB} = -1$''. In cases VII and VIII the $z_{\rm BB}
= \infty$ is created by the first term on the right-hand side of (\ref{6.9}),
and the behaviour of the second term is irrelevant. In the other cases, whether
$z_{\rm BB} = -1$ is possible or not depends on the limit at the BB of
$\left[(\Phi,_r/\Phi) (k^r)^2 {\cal D}_k / J^2\right]$ inside $L(s)$, and also
on the sign of ${\cal D}_k$ (depending on this sign, the limit, if infinite, may
be $+ \infty$ or $- \infty$). Case V becomes an infinitely blueshifted radial
ray in the L--T limit.

\section{Null geodesics in symmetric subcases of the Szekeres
spacetimes}\label{symmetric}

\setcounter{equation}{0}

In special cases, first integrals of the geodesic equations (\ref{5.3}) --
(\ref{5.6}) exist. One of them is when $Q$ is constant; then there exist null
geodesics along which $y = Q$ and $k^y = 0$.\footnote{Note that the case when $P
=$ constant and $x = P$ along the geodesic is equivalent to $Q =$ constant and
$y = Q$ under the transformation $(x, y) = (y', x')$ accompanied by the
relabeling $(P, Q) = (\widetilde{Q}, \widetilde{P})$, which does not change the
metric.} Equation (\ref{5.6}) is then fulfilled identically.

\subsection{Null geodesics in the axially symmetric subcase}

Another special case is when $P$ and $Q$ are constant \cite{BKHC2010,NoDe2007}.
Then the Szekeres spacetime is axially symmetric around $(x, y) = (P, Q)$, and a
family of null geodesics exists on which $x = P$ and $y = Q$. The transformation
\begin{equation}\label{7.1}
x = x' + P, \qquad y = y' +Q
\end{equation}
has then the same result as if
\begin{equation}\label{7.2}
P = Q = 0,
\end{equation}
which we shall assume. Then we introduce
\begin{equation}\label{7.3}
x' = u \cos \varphi, \qquad y' = u \sin \varphi,
\end{equation}
which changes (\ref{2.1}) and (\ref{2.2}) to
\begin{equation}\label{7.4}
{\rm d} s^2 = {\rm d} t^2 - \frac {{\cal N}^2 {\rm d} r^2} {1 + 2 E(r)} -
\left(\frac {\Phi} {\cal E}\right)^2 \left({\rm d} u^2 + u^2 {\rm d}
\varphi^2\right),\ \ \ \ \
\end{equation}
\begin{equation}\label{7.5}
{\cal E} = \frac 1 {2S}\ \left(u^2 + S^2\right);
\end{equation}
the dipole equator ${\cal E},_r = 0$ is now at $u = S$. In these coordinates,
the geodesic equations for (\ref{7.4}) -- (\ref{7.5}) are
\begin{eqnarray}
\dr {k^t} {\lambda} &+& \frac {{\cal N} {\cal N},_t} {1 + 2E} \left(k^r
\right)^2 + \frac{\Phi {\Phi,_t}}{{\cal E}^2} \left[\left(k^u\right)^2 + u^2
\left(k^{\varphi}\right)^2\right] = 0, \nonumber \\
\label{7.6} \\
\dr {k^r} {\lambda} &+& 2 \frac {{\cal N},_t} {\cal N} k^t k^r \nonumber \\
&+& \left(\frac {{\cal N},_r} {\cal N} - \frac {E,_r} {1 + 2E}\right)
\left(k^r\right)^2 + 2 \frac {u \Phi S,_r} {S {\cal E}^2 {\cal N}}\ k^r k^u
\nonumber \\
&-& \frac {\Phi} {{\cal E}^2} \frac {1 + 2E} {{\cal N}} \left[\left(k^u\right)^2
+ u^2 \left(k^{\varphi}\right)^2\right] = 0, \label{7.7}
\end{eqnarray}
\begin{eqnarray}
\dr {k^u} {\lambda} &+& 2 \frac {\Phi,_t} {\Phi} k^t k^u - \frac {u S,_r {\cal
N}} {S \Phi (1 + 2E)}\ \left(k^r\right)^2 + 2 \frac {\cal N} {\Phi} k^r k^u
\nonumber \\
&-& \frac u {S {\cal E}}\ \left(k^u\right)^2 + u \left(\frac {u^2} {S {\cal E}}
- 1\right) \left(k^{\varphi}\right)^2 = 0, \label{7.8} \\
\dr {k^{\varphi}} {\lambda} &+& 2 \frac {\Phi,_t} {\Phi} k^t k^{\varphi} + 2
\frac {\cal N} {\Phi} k^r k^{\varphi} + 2 \left(\frac 1 u - \frac u {S {\cal
E}}\right) k^u k^{\varphi} = 0. \nonumber \\ \label{7.9}
\end{eqnarray}
The remark made at the end of Sec. \ref{nullgeo} applies also here: the
coefficients of $(k^r)^2$ and of $k^r$ in (\ref{7.8}) and (\ref{7.9}) require
special treatment. However, these equations become regular when the geodesic
stays within the $\{\varphi = {\rm constant}, u = 0\}$ surface. Such a geodesic
intersects every space of constant $t$ on the symmetry axis.

Equation (\ref{7.9}) has the first integral:
\begin{equation}\label{7.10}
k^{\varphi} u^2 \Phi^2 / {\cal E}^2 = J_0,
\end{equation}
where $J_0$ is constant along each geodesic (not necessarily null). When
(\ref{7.10}) is substituted in (\ref{5.7}) transformed to the $(u, \varphi)$
coordinates, the following results:
\begin{equation}\label{7.11}
(k^t)^2 = \frac {{\cal N}^2 \left(k^r\right)^2} {1 + 2E} + \left(\frac
{\Phi} {\cal E}\right)^2 \left(k^u\right)^2 + \left(\frac {J_0 {\cal E}} {u
\Phi}\right)^2.
\end{equation}
At the observation point (\ref{5.8}) applies, at the emission point (\ref{6.2})
can be used.

Equations (\ref{7.11}) and (\ref{6.2}) show that for geodesics emitted at the
BB, where $\Phi = 0$, the observed redshift is infinite when $J_0 \neq 0$. A
necessary (but not sufficient) condition for $1 + z_o = 0$ is $J_0 = 0$, i.e.
the ray must proceed within the hypersurface of constant $\varphi$.

\subsection{Null geodesics in the L--T limit}

The L--T model follows from (\ref{2.1}) -- (\ref{2.2}) as the limit $P,_r = Q,_r
= S,_r = 0$ ($\Longrightarrow {\cal E},_r = 0$). Then the spheres of constant
$t$ and $r$ become concentric, the spacetime becomes spherically symmetric and
(\ref{5.5}) -- (\ref{5.6}) imply
\begin{equation}\label{7.12}
\left(k^x\right)^2 + \left(k^y\right)^2 = \frac {C^2 {\cal E}^2} {\Phi^4},
\end{equation}
where $C$ is constant along the geodesic. Thus, the geodesic is radial ($k^x =
k^y = 0$ along it) when $C = 0$.

When substituted in (\ref{6.3}), Eq. (\ref{7.12}) implies that along all
nonradial rays $z \to \infty$ at the BB (where $\Phi \to 0$) irrespectively of
whether $\dril {t_B} r = 0$ or not \cite{Kras2016}.

\section{The Extremum Redshift Surface}\label{ERS}

\setcounter{equation}{0}

In the L--T limit, an Extremum Redshift {\it Hyper}surface (ERH) was defined. It
is the locus where the redshift along past-directed {\it radial} rays achieves a
local maximum or minimum. Since in L--T models the collection of all radial rays
at any fixed time is two-parametric, the ERH is a 3-dimensional hypersurface in
spacetime. But in a general Szekeres spacetime, radial directions are not
defined. In the axisymmetric subcase, an analogue of the ERH exists, but, as
will be seen below, it is 2-dimensional, so it will be called the Extremum
Redshift {\it Surface}.

Consider a null geodesic that stays in the surface $\{u, \varphi\} = \{0, {\rm
constant}\}$; it obeys (\ref{7.8}) and (\ref{7.9}) identically. The remark made
under (\ref{5.7}) applies to it in an even stronger form: $k^r \neq 0$ at all
points because with $u = 0 = k^u = k^{\varphi}$ the geodesic would be timelike
wherever $k^r = 0$. Assume the geodesic is past-directed and has its initial
point at $r = 0$. Thus, $r$ has to increase on it and can be used as a
parameter. Using (\ref{6.2}), we rewrite (\ref{7.6}) as follows:
\begin{equation}\label{8.1}
\dr z {\lambda} = \frac {{\cal N} {\cal N},_t} {1 + 2E}\ k^r \dr r {\lambda}.
\end{equation}
Changing the parameter to $r$ we obtain
\begin{equation}\label{8.2}
\dr z r = \frac {{\cal N} {\cal N},_t} {1 + 2E}\ k^r.
\end{equation}
Since ${\cal N} \neq 0$ from no-shell-crossing conditions \cite{HeKr2002} and
$k^r > 0$, the extrema of $z$ on such a geodesic occur where
\begin{equation}\label{8.3}
{\cal N},_t \equiv \Phi,_{tr} - \Phi,_t {\cal E},_r/{\cal E} = 0.
\end{equation}
Equation (\ref{8.3}) was derived under the assumption $u = 0$. Thus, the set in
spacetime defined by (\ref{8.3}) is 2-dimensional ($\varphi$ is constant, but
arbitrary). We will call it Extremum Redshift Surface (ERS).

{}From (\ref{2.4}) and (\ref{3.2}) with $\Lambda = 0$ we obtain
\begin{equation}\label{8.4}
\Phi,_t = r \sqrt{\frac {2 M_0 r} {\Phi} - k},
\end{equation}
{}From now on we proceed assuming the function $E(r)$ in the same form as in
Ref. \cite{Kras2016},
\begin{equation}\label{8.5}
2 E(r) = - kr^2, \quad {\rm where} \quad -k = 0.4
\end{equation}
(see Appendix \ref{genERS} for the ERS equation without this simplification).
Then $(3/2)E,_r/E - M,_r/M = 0$, and from (\ref{4.2}) we obtain
\begin{equation}\label{8.6}
\Phi,_{tr} = \sqrt{\frac {2 M_0 r} {\Phi} - k} + \frac {M_0 r^3} {\Phi^2}\
t_{B,r}.
\end{equation}
Using this, (\ref{8.4}) and (\ref{7.5}) with $u = 0$, Eq. (\ref{8.3}) becomes
\begin{equation}\label{8.7}
\sqrt{\frac {2 M_0 r} {\Phi} - k} \left(1 - r \frac {S,_r} S\right) = - \frac
{M_0 r^3} {\Phi^2}\ t_{B,r}.
\end{equation}

\noindent To avoid shell crossings, $t_{B,r} < 0$ must hold \cite{HeKr2002}, so
the right-hand side of (\ref{8.7}) is positive. The left-hand side is positive
in consequence of (\ref{2.8}) and (\ref{3.2}).

With (\ref{8.5}), $E > 0$ and we can use (\ref{a.3}) for $\Phi$. Squaring both
sides of (\ref{8.7}) and remembering that $k < 0$ we obtain
\begin{equation}\label{8.8}
(\cosh \eta + 1) (\cosh \eta - 1)^3 = - \frac {k^3 r^2 {t_{B,r}}^2} {{M_0}^2
\left(1 - r S,_r / S\right)^2}.
\end{equation}
Denoting
\begin{equation}\label{8.9}
\chi \df \sinh^2 (\eta/2)
\end{equation}
Eq. (\ref{8.8}) can be written as
\begin{equation}\label{8.10}
\chi^4 + \chi^3 = - k^3 \left[\frac {r t_{B,r}} {4 M_0 \left(1 - r S,_r /
S\right)}\right]^2.
\end{equation}
The ERH equation in Ref. \cite{Kras2014} follows from (\ref{8.10}) as the limit
$S,_r = 0$. With $k < 0$, (\ref{8.10}) is solvable for $\chi$ at any $r$, since
its left-hand side is independent of $r$ and can vary from 0 to $+\infty$ while
the right-hand side is positive.

\section{An exemplary axially symmetric QSS model}\label{exQSS}

\setcounter{equation}{0}

Since, so far, it turned out to be impossible to determine the rays with
infinite blueshift by exact calculations, we shall now attempt to detect them
numerically in the Szekeres spacetimes given by (\ref{7.4}) -- (\ref{7.5}). In
choosing a simple form for $S(r)$ one must take care to obey (\ref{2.6}) and
(\ref{2.7}), which, using (\ref{3.2}) and (\ref{8.5}), imply
\begin{equation}\label{9.1}
1/r > S,_r / S.
\end{equation}
Equations (\ref{9.1}) and (\ref{3.4}) will be fulfilled when
\begin{equation}\label{9.2}
S = \sqrt{a^2 + r^2},
\end{equation}
where $a > 0$ is a constant. With (\ref{7.2}), the equation of the dipole
``equator'' ${\cal E},_r = 0$ becomes
\begin{equation}\label{9.3}
x^2 + y^2 = S^2,
\end{equation}
and the axis of symmetry is $x = y = 0$.

To define a model completely we need to prescribe the bang-time function
$t_B(r)$. We choose it in the form
\begin{equation}\label{9.4}
t_B(r) = \left\{\begin{array}{ll}
 A \left({\rm e}^{- \alpha r^2} - {\rm e}^{- \alpha {r_b}^2}\right) + t_{\rm
 BB} & \quad {\rm for}\ r \leq r_b, \\
 t_{\rm BB} & \quad {\rm for}\ r \geq r_b, \\
 \end{array} \right.
\end{equation}
where $A, \alpha, r_b$ and $t_{\rm BB}$ are constants. For $r \geq r_b$ this
spacetime goes over into the Friedmann spacetime (see Appendix \ref{frlim}),
albeit represented in exotic coordinates.

\begin{figure}[h]
\hspace{-5mm} \includegraphics[scale=0.6]{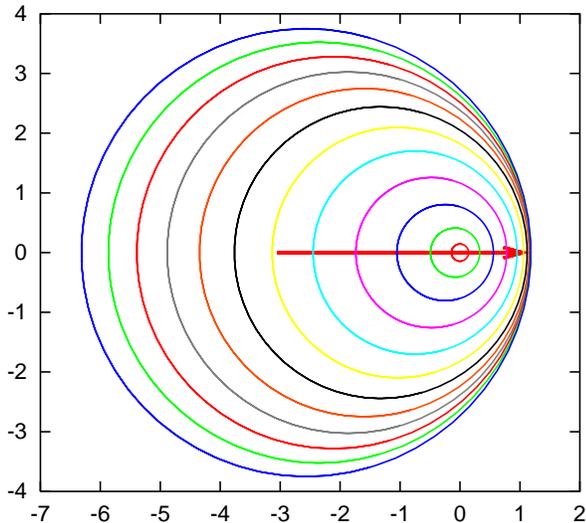} \caption{The
constant-$r$ circles in the $(r, u)$ surface mapped into a plane so that the
smallest distances between them are the same as in the metric (\ref{7.4}). The
arrow points along the direction of the dipole maxima, $u = 0$. See text for
more explanations.} \label{circles}
\end{figure}

Figure \ref{circles} shows the cross-section of the spacetime by a surface of
(any) constant $\varphi$ and $t = t_o = 1.2$ (most rays considered further on
will have their initial points at this $t$). Each such surface consists of
non-concentric circles, but is not flat, so Fig. \ref{circles} is not an
isometric image.

A definition of the radius of each circle is not self-evident. From (\ref{2.1})
it is seen that the radius can be defined either as the curvature radius
$\Phi(t_o, r)$ or as a geodesic radius, by integrating $\sqrt{|g_{rr}|}$. But
the value of the integral depends on the direction in the $(x, y)$ surface, and
the centers of circles of different radii do not coincide. The most natural
definition seems to be
\begin{equation}\label{9.5}
{\cal R}(r_f) \df \int_0^{r_f} \frac {\Phi,_r(t_o, r)} {\sqrt{1 + 2E(r)}}\ {\rm
d} r
\end{equation}
for two reasons:

(1) This path of integration goes along the dipole equator ${\cal E},_r = 0$, so
such ${\cal R}$ is the same as in an L--T model.

(2) ${\cal R}$ as defined by (\ref{9.5}) coincides with
\begin{equation}\label{9.6}
\widetilde {\cal R} \df \int_0^{r_f} \frac {\Phi,_r(t_o, r) - \Phi(t_o, r) {\cal
E},_r / {\cal E}} {\sqrt{1 + 2E(r)}}\ {\rm d} r
\end{equation}
averaged over all directions, i.e. over the whole sphere $r = r_f$. From
(\ref{2.1}), the surface element of such a sphere is $\left[\Phi(t_o, r_f)/{\cal
E}(r_f, x, y)\right]^2 {\rm d} x {\rm d} y$, its surface area is $4 \pi
\Phi^2(t_o, r_f)$, and so the average of $\widetilde{\cal R}$ over this sphere
is
\begin{equation}\label{9.7}
\langle\widetilde{\cal R}\rangle = {\cal R}
\end{equation}
because
\begin{eqnarray}
\int_{-\infty}^{+\infty} {\rm d} x \int_{-\infty}^{+\infty} {\rm d} y\ \frac
{{\cal E},_r} {{\cal E}^3} &=& 0, \label{9.8} \\
\int_{-\infty}^{+\infty} {\rm d} x \int_{-\infty}^{+\infty} {\rm d} y\ \frac 1
{{\cal E}^2} &=& 4 \pi. \label{9.9}
\end{eqnarray}
Equations (\ref{9.7}) -- (\ref{9.9}) apply with any ${\cal E}$ given by (\ref{2.2}).

The $r$-coordinates of the circles in Fig. \ref{circles} run from $r_0 = 0$ to
$r_{12} = 2.4$ at intervals of $\Delta r = 0.2$. Their radii were calculated
from (\ref{9.5}) with $M$, $E$ and ${\cal E}$ given by (\ref{3.2}), (\ref{8.5})
and (\ref{7.5}), respectively. The distances between them were calculated along
the dipole maximum ($u = 0$), from
\begin{eqnarray}\label{9.10}
&& d_{\rm max} = \int_{r_i}^{r_{i + 1}} \frac {\Phi,_r(t_o, r) - \Phi(t_o, r)
S,_r / S} {\sqrt{1 + 2E(r)}}\ {\rm d} r, \nonumber \\
&& i = 0, 1, \dots, 12,
\end{eqnarray}
because $\left.({\cal E},_r/{\cal E})\right|_{u = 0} = S,_r/S$. The three
largest circles are in the Friedmann region.\footnote{In the Friedmann region,
the circles are non-concentric in consequence of the coordinate choice, see
Appendix \ref{frlim}.}

Figure \ref{circles} is drawn so that the shortest distances between the circles
are the same as the $d_{\rm max}$ in (\ref{9.10}). If the circles were drawn so
that the longest distances between them were the same as along the dipole
minimum in the curved surface (i.e. with + in the numerator of (\ref{9.10})),
the image would be the same.

\section{Axial rays in the axially symmetric QSS model}\label{simpleblue}

\setcounter{equation}{0}

\begin{figure}[h]
\hspace{-5mm} \includegraphics[scale=0.7]{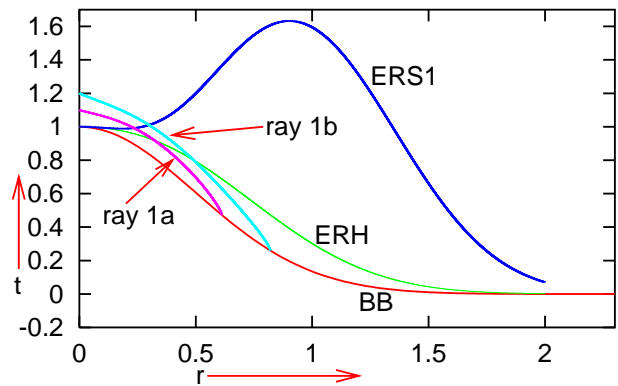} \caption{The $t(r)$
graphs of two rays approaching the Big Bang given by (\ref{9.4}) within the
surface $\{u, \varphi\} = \{0, {\rm constant}\}$ in the metric (\ref{7.4}) --
(\ref{7.5}). See text for more explanation.} \label{axcones1}
\end{figure}

For the numerical examples we chose
\begin{equation}\label{10.1}
(A, \alpha, t_{\rm BB}, r_b) = (1, 2, 0, 2).
\end{equation}
The first calculation was for two null geodesics going back in time with $u = 0
= k^{\varphi}$ from $(r, t) = (0, t_B(0) + 0.1)$ and $(r, t) = (0, t_B(0) +
0.2)$, respectively, in the model with $a^2 = 0.1$, see Fig. \ref{axcones1}. The
numerical calculation confirmed that $z \to -1$ as the rays approach the BB, see
Table \ref{table1}. This model belongs to case V in Table \ref{treetable}.

Figure \ref{axcones1} shows the BB profile, the $t(r)$ graphs of the two rays
and the ERS profile corresponding to $a^2 = 0.1$. The ERH profile in the L--T
spacetime with the same $t_B(r)$ and $E(r)$ is also included. It can be seen
that in a Szekeres model, the rays enter the ERS (i.e. begin acquiring negative
contributions to redshift) at smaller $r$ than in the corresponding L--T model.
Note that, in consequence of the mass dipole, the geometrical distance between
$r_1$ and $r_2$ along a line of constant $(t, u, \varphi)$ and ${\cal E},_r > 0$
is shorter in a Szekeres spacetime than in the corresponding L--T spacetime.
Thus, the figure is not a faithful image of the geometrical relations.

The jump in the ERS profile at $r = 2$ is a consequence of the jump in $\dril
{t_B} r$ at $r = r_b = 2$; see (\ref{9.4}) and (\ref{8.10}). Similar jumps will
be seen in other ERS profiles in the next figures. The ERH in Fig.
\ref{axcones1} also has a jump at $r = 2$, but, with the parameter values given
by (\ref{10.1}) and $a^2 = 0.1$, the right-hand side of (\ref{8.10}) is 1681
times smaller in an L--T model than in the Szekeres model, so the jump in $\chi$
is also much smaller.

\begin{figure}[h]
${ }$ \\ [5mm]
\includegraphics[scale=0.6]{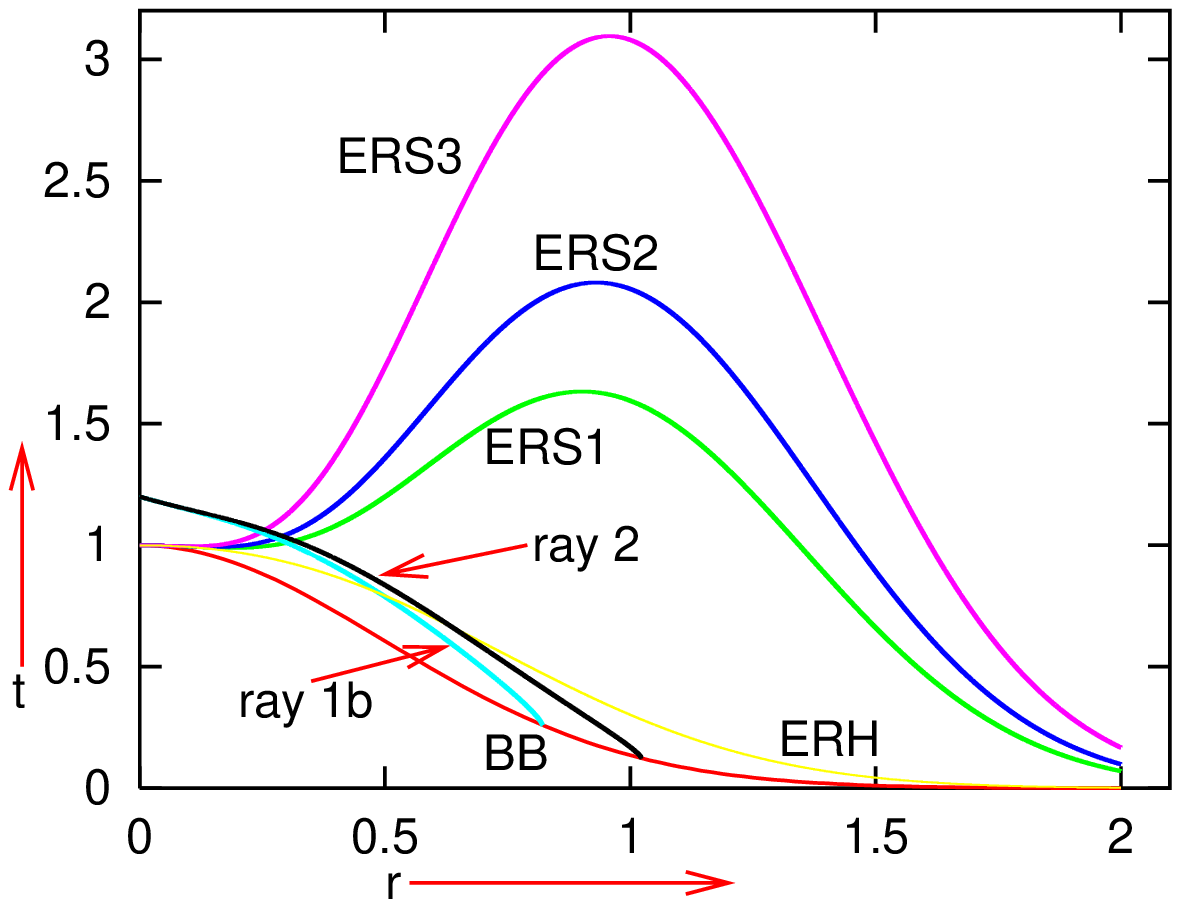}
${ }$ \\ [5mm]
\includegraphics[scale=0.6]{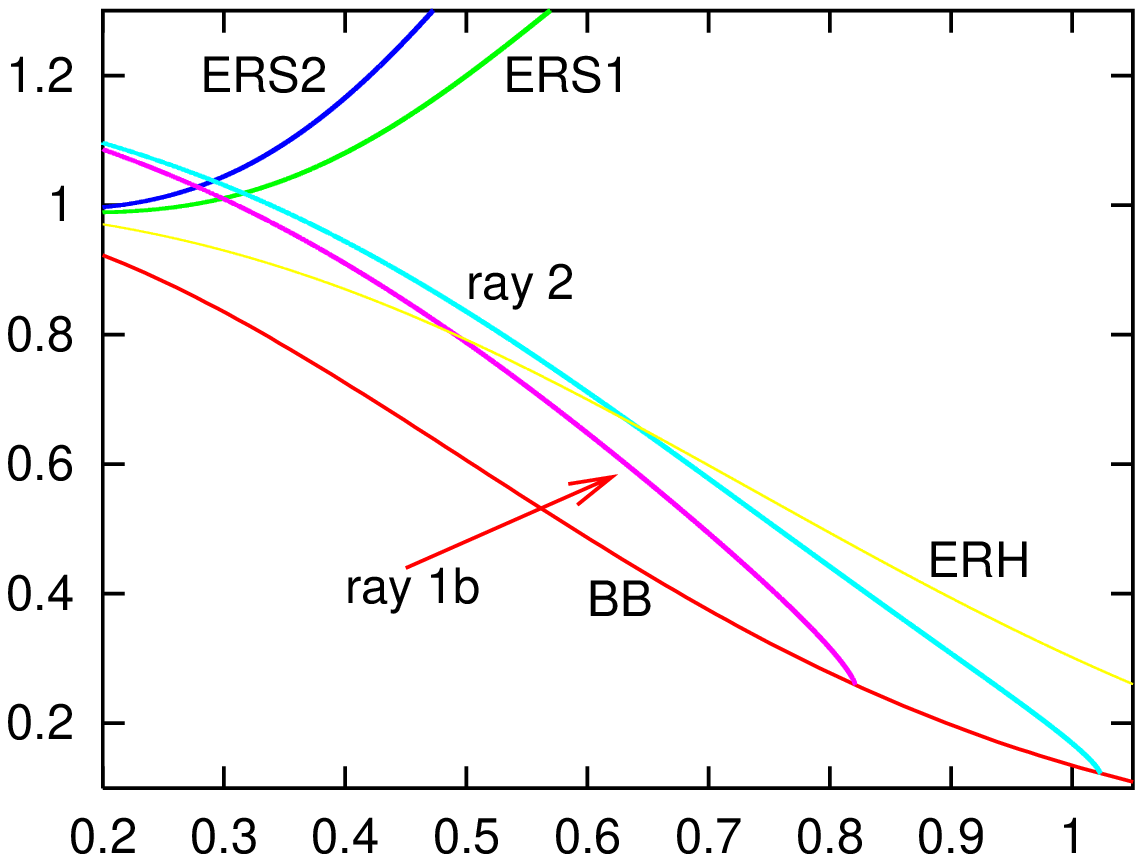}
\caption{{\bf Upper panel:} Smaller value of $a$ results in a higher maximum of
the ERS. See text for more explanation. {\bf Lower panel:} Closeup view on the
rays; the ERS3 is omitted.} \label{axcones2}
\end{figure}

Figure \ref{axcones2} demonstrates the influence of the value of $a$ on the
shape of the ERS. Ray 1b and ERS1 are the same as in Fig. \ref{axcones1}. Ray 2
and ERS2 were calculated with $a^2 = 0.07$, ERS3 was calculated with $a^2 =
0.04$; the values of the other parameters were the same as for ray 1b. As seen,
smaller $a$ gives a stronger effect. Larger $a$ produces a smaller difference
between the ERS and the ERH of the corresponding L--T model. In the limit $a \to
\infty$ the L--T result would be recovered -- as can be seen from (\ref{9.2})
and (\ref{8.10}), this limit has the same effect as $S,_r = 0$. The added
flexibility in the Szekeres models is that the time of flight of the ray under
the ERS can be increased by manipulating the $S$ function. In the L--T models,
increasing this time was possible only by manipulating the BB profile. The lower
panel of Fig. \ref{axcones2} is a closeup view on the region where many lines
intersect.

\begin{table*}
\caption{Parameters of the rays from Fig. \ref{axcones1}}
\bigskip
\begin{tabular}{|l|l|l|}
  \hline \hline
\ \ Parameter\ \ & \ \ Ray 1a (lower) \ \ & \ \ Ray 1b (upper)\ \ \\
  \hline \hline
 \ \ $t$ at $r = 0$ & \ \ $t_B(0) + 0.1$ & \ \ $t_B(0) + 0.2$ \\
  \hline
 \ \ $r$ at the BB & \ \ 0.61238227292746328 & \ \ 0.82090257143313361 \\
  \hline
 \ \ $t$ at the BB & \ \ 0.47201970792080822 & \ \ 0.25948561090989042 \\
\hline
 \ \ $1 + z$ at the BB & \ \ $2.53722758372994317 \times 10^{-6}$ \ \ & \ \
 $2.63658970817360853 \times 10^{-6}$ \ \ \\
  \hline
 \ \ maximum $z$ & \ \ 0.44409843877390864 & \ \ 0.44862603680491642 \\
 \hline
 \ \ $r$ at maximum $z$\ \ & \ \ 0.23503329565169392 & \ \ 0.29909096252077694 \\
 \hline
 \ \ $t$ at maximum $z$ & \ \ 0.99199314244492787 & \ \ 1.0102113079060080 \\
 \hline
  \ \ $r$ at $z = 0$ & \ \ r = 0.46613259040713573 & \ \ 0.61839965892230220 \\
 \hline
  \ \ $t$ at $z = 0$ & \ \ 0.74684282247031808 & \ \ 0.62045744911301359 \\
  \hline \hline
\end{tabular}
\label{table1}
\end{table*}

\begin{figure}[h]
\includegraphics[scale=0.6]{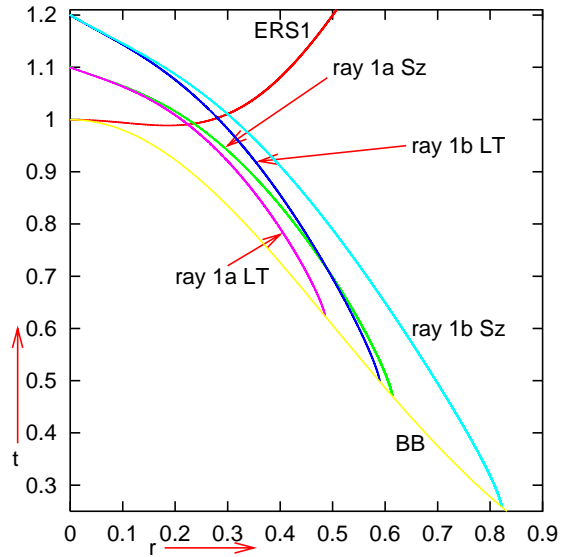}
\caption{Rays 1a and 1b from Fig. \ref{axcones1} compared with their L--T
counterparts. See text for more explanation.} \label{comprays}
\end{figure}

\begin{figure}[h]
\includegraphics[scale=0.64]{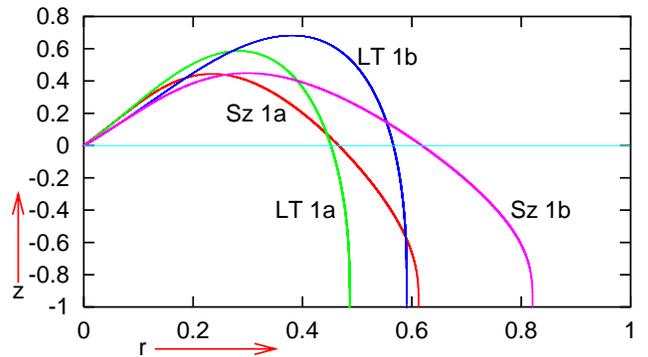}
\caption{Redshift profiles along the rays from Fig. \ref{axcones1} compared with
the corresponding redshift profiles in the L--T model. See text for more
explanation. At the intersections of the rays with the Big Bang all redshifts
approach $-1$.} \label{compreds}
\end{figure}

Figure \ref{comprays} shows the comparison of rays 1a and 1b from Fig.
\ref{axcones1} with the rays in the L--T model that have the same initial $(r,
t)$; the BB profile is the same in both models. The rays in the Szekeres model
hit the BB at larger values of $r$. In the L--T model, increasing the $r$
coordinate of the intersection of the ray with the BB required moving up the
initial point of the ray, in the Szekeres model this can be achieved by
increasing $S,_r/S$ without changing the initial data of the ray.

Figure \ref{compreds} shows the redshift profiles along the rays seen in Fig.
\ref{comprays}. The maximum redshift along each Szekeres ray is smaller than the
maximum along the corresponding L--T ray. This is consistent with the message of
Fig. \ref{axcones1}: the Szekeres rays begin acquiring negative contributions to
redshift at smaller $r$, so the maximum $z$ is smaller for the Szekeres rays
than for the corresponding L--T rays.

\section{Other rays with blueshift in the axially symmetric case}\label{moreblue}

\setcounter{equation}{0}

As seen from (\ref{7.11}), a necessary condition for the existence of blueshift
along a ray emitted at the BB is $J_0 = 0$, i.e. the ray must proceed within a
hypersurface of constant $\varphi$. In this section, we will maintain this
assumption, but will relax the assumption $u = 0$ along the ray that was adopted
in Sec. \ref{simpleblue}. Then (\ref{7.9}) is still fulfilled identically. As
remarked at the end of Sec. \ref{nullgeo}, the coefficients ${\cal N}/\Phi$ in
front of $(k^r)^2$ and $k^r$ in (\ref{7.8}) become infinite at $r = 0$.
Therefore, the limits at $r \to 0$ of these whole terms have to be calculated
exactly. In the calculation below we assume that at $r = 0$ both $k^r$ and $k^u$
are finite and that the calculation is done at $t > t_B$.

With $S$ chosen as in (\ref{9.2}) we have $\lim_{r \to 0} S,_r = 0$ and $S(0) =
a > 0$. Knowing this we obtain using (\ref{4.11})
\begin{equation}\label{11.1}
\lim_{r \to 0} \frac {S,_r} {\Phi} = \frac 1 {a R} < \infty,
\end{equation}
so the third term in (\ref{7.8}) is finite at $r = 0$.

For calculating the limit at $r \to 0$ of the fourth term in (\ref{7.8}) we
observe that it can be finite only when either $\lim_{r \to 0} k^r = 0$ or
$\lim_{r \to 0} k^u = 0$. However, with $J_0 = 0 = k^r(0)$, Eq. (\ref{7.11})
becomes a contradiction with (\ref{5.8}) at $r = 0$, so $\lim_{r \to 0} k^r \neq
0$. Consequently,
\begin{equation}\label{11.2}
\lim_{r \to 0} k^u = 0.
\end{equation}

We now calculate $\lim_{r \to 0} (k^u/\Phi)$ along a null geodesic. We begin
with the de l'H{\^o}pital rule and use (\ref{7.8}) for $\dril {k^u} {\lambda}$.
Then we use $k^{\varphi} = 0$ and (\ref{11.1}). The result is
\begin{eqnarray}\label{11.3}
&& \lim_{r \to 0} \frac {k^u} {\Phi} = \lim_{r \to 0} \frac {\dril {k^u}
{\lambda}} {\Phi,_t k^t + \Phi,_r k^r} = \nonumber \\
&&- \lim_{r \to 0} \left\{\frac 1 {\Phi,_r k^r} \left[\frac {\Phi,_t} {\Phi}\
k^t k^u - \frac {u {\cal N}} {a R S (1 + 2E)}\ (k^r)^2\right.\right. \nonumber
\\
&&\ \ \ \ \ + \left.\left.2 {\cal N} k^r \frac {k^u} {\Phi} - \frac {u (k^u)^2}
{S {\cal E}}\right]\right\}.
\end{eqnarray}
In this, we use (\ref{4.10}), (\ref{11.2}), (\ref{7.11}) with $J_0 = 0$ taken at
the observation point where $k^t = -1$ and $\Phi = 0$, (\ref{5.2}),
(\ref{4.11}), (\ref{8.5}) and (\ref{9.2}). Solving the result for $\lim_{r \to
0} (k^u/\Phi)$ we obtain
\begin{equation}\label{11.4}
\lim_{r \to 0} \frac {k^u} {\Phi} = \frac u {3 a^2 R^2},
\end{equation}
so the fourth term in (\ref{7.8}) is also finite at the origin.

Since numerical calculations cannot respect such intricate limits automatically,
the initial points for geodesics passing through the origin must be chosen at
the origin, as remarked at the end of Sec. \ref{nullgeo}.

For the next experiments, we take (past-directed) null geodesics with
$k^{\varphi} \equiv 0$ and initial $u_o > 0$. Note, from (\ref{7.8}), that if
$u_o \neq 0$ then $\left(\dril {k^u} {\lambda}\right)_o \neq 0$ even if $k_o^u =
0$, so the ray will not stay in a constant-$u$ surface. We consider the
following cases:
 \begin{itemize}
\item (I) $u_o = S/100$. This is close to the case $u = 0$ investigated in Sec.
\ref{simpleblue}, where ${\cal E},_r/{\cal E}$ had a maximum equal to $S,_r/S$.
\item (II) $u_o = S/10$.
\item (III) $u_o = S/2$.
\item (IV) $u_o = S$, where the contribution of the dipole is zero, ${\cal
E},_r = 0$.
\item (V) $u_o = 2S$.
\item (VI) $u_o = 10S$.
\item (VII) $u_o = 100S$.
\item (VIII) $u_o \to \infty$, where the dipole contribution has the minimum
equal to $(- S,_r/S)$.
 \end{itemize}
For $u > S$ the metric and the Christoffel symbols will have to be transformed
to the new coordinate $w = 1/u$. This is because with $u$ becoming very large,
$k^u$ tends to $\infty$ much faster than $u$ and stops the program before the
calculation comes near to the BB.

In every case, the $(t, r)$ coordinates of the initial point will be the same as
for ray 1b in the previous examples,
\begin{equation}\label{11.5}
(t, r)_o = (t_B(0) + 0.2,\ 0).
\end{equation}
At the initial point, where $\Phi = 0$, we use (\ref{7.11}) with $J_0 = 0$,
(\ref{5.8}), (\ref{3.6}) -- (\ref{3.7}), (\ref{3.2}), (\ref{8.5}) and
(\ref{a.3}) to find
\begin{equation}\label{11.6}
k_o^r = \frac {-k} {M_0 (\cosh \eta_o - 1)},
\end{equation}
where $\eta_o$ is at $t = t_B(0) + 0.2$. Next, using (\ref{11.4}), (\ref{11.1}),
(\ref{4.11}), (\ref{8.5}) and (\ref{11.6}) we obtain from (\ref{7.8})
\begin{equation}\label{11.7}
\left.\dr {k^u} {\lambda}\right|_o = \left.\frac {u k^2} {3 a S {M_0}^2 (\cosh
\eta - 1)^2}\right|_o.
\end{equation}

In the transformed coordinate $w = 1/u$, the metric and the function ${\cal N}$
are replaced by
\begin{eqnarray}
&& {\rm d} s^2 = {\rm d} t^2 - \frac {{\widetilde{\cal N}}^2 {\rm d} r^2} {1 + 2
E(r)} - \left(\frac {\Phi} {\widetilde{\cal E}}\right)^2 \left({\rm d} w^2 + w^2
{\rm d} \varphi^2\right), \nonumber \\
\label{11.8} \\
&& \widetilde{\cal N} \df \Phi,_r - \Phi \widetilde{\cal E},_r/\widetilde{\cal
E}, \label{11.9}
\end{eqnarray}
where
\begin{equation}\label{11.10}
\widetilde{\cal E} = \frac {w^2 S} 2 + \frac 1 {2S},
\end{equation}
and the equator of the dipole is at $w = 1/S$. The only changes in (\ref{7.6}),
(\ref{7.10}) and (\ref{7.11}) are $(u, {\cal N}, {\cal E}) \to (w,
\widetilde{\cal N}, \widetilde{\cal E})$), while (\ref{7.7}) -- (\ref{7.9})
change to
\begin{eqnarray}
\dr {k^r} {\lambda} &+& 2 \frac {{\widetilde{\cal N}},_t} {\widetilde{\cal N}}
k^t k^r \nonumber \\
&+& \left(\frac {{\widetilde{\cal N}},_r} {\widetilde{\cal N}} - \frac {E,_r} {1
+ 2E}\right) \left(k^r\right)^2 - 2 \frac {w \Phi S,_r} {S {\widetilde{\cal
E}}^2 {\widetilde{\cal N}}}\ k^r k^w \nonumber \\
&-& \frac {\Phi} {{\widetilde{\cal E}}^2} \frac {1 + 2E} {{\widetilde{\cal N}}}
\left[\left(k^w\right)^2 + w^2 \left(k^{\varphi}\right)^2\right] = 0, \label{11.11} \\
\dr {k^w} {\lambda} &+& 2 \frac {\Phi,_t} {\Phi} k^t k^w + \frac {w S,_r
{\widetilde{\cal N}}} {S \Phi (1 + 2E)}\ \left(k^r\right)^2 + 2 \frac
{\widetilde{\cal N}} {\Phi} k^r k^w \nonumber \\
&-& \frac {w S} {\widetilde{\cal E}}\ \left(k^w\right)^2 + w \left(\frac {w^2 S}
{\widetilde{\cal E}} - 1\right) \left(k^{\varphi}\right)^2 = 0, \label{11.12} \\
\dr {k^{\varphi}} {\lambda} &+& 2 \frac {\Phi,_t} {\Phi} k^t k^{\varphi} + 2
\frac {\widetilde{\cal N}} {\Phi} k^r k^{\varphi} + 2 \left(\frac 1 w - \frac {w
S} {\widetilde{\cal E}}\right) k^w k^{\varphi} = 0. \nonumber \\ \label{11.13}
\end{eqnarray}
Also, (\ref{11.4}) and (\ref{11.7}) change to:
\begin{eqnarray}
\lim_{r \to 0} \frac {k^w} {\Phi} &=& - \frac w {3 a^2 R^2}, \label{11.14} \\
\left.\dr {k^w} {\lambda}\right|_o &=& - \left.\frac {w k^2} {3 a S {M_0}^2
(\cosh \eta - 1)^2}\right|_o. \label{11.15}
\end{eqnarray}

\begin{figure}[h]
\hspace{-5mm} \includegraphics[scale=0.64]{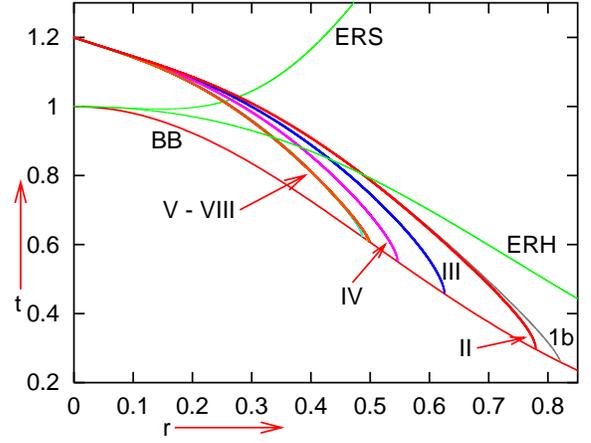} \caption{The $t(r)$
graphs of rays II -- VIII compared with the graph for ray 1b. Ray I nearly
coincides with ray 1b and is omitted. Ray IV has its initial point at the
equator of the dipole. See text for more explanation.} \label{axsymrays}
\end{figure}

Figure \ref{axsymrays} shows the $t(r)$ graphs of rays I -- VIII compared with
the $t(r)$ graph of ray 1b. Ray I very nearly coincides with ray 1b. As $u_o$
increases, the ray hits the BB at ever smaller $r$. With $u_o > S$ the graphs
become ever closer to each other. The lower end of ray V is barely visible at
the BB, rays VI -- VIII coincide at the scale of the figure. Ray VIII, similarly
to ray 1b, proceeds within the surface of constant $\varphi$ and $u$, but with
$u = \infty$ ($w = 0$).

Only rays 1b and VIII run within the $(t, r)$ surface shown in Fig.
\ref{axsymrays}. Along the other rays $u$ varies, so this figure shows the
projections of those rays on the $(t, r)$ surface along lines of constant $(t,
r)$.

\begin{figure}[h]
\hspace{-5mm} \includegraphics[scale=0.64]{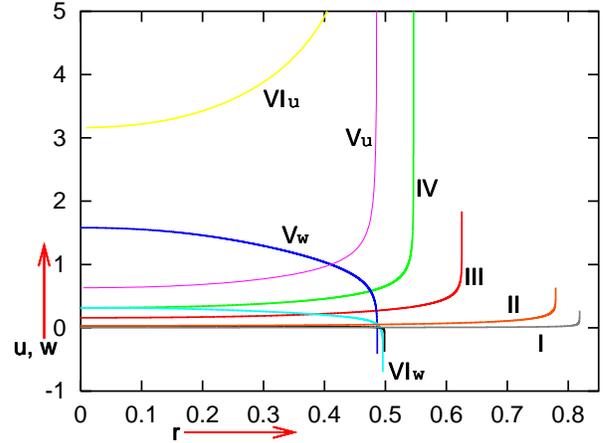} \caption{The $u(r)$
and $w(r)$ graphs for rays I -- VII. See text for more explanation.}
\label{urgraphs}
\end{figure}

Figure \ref{urgraphs} shows the projections of rays I -- VII on the $(u, r)$
surface along the lines of constant $u$ and $r$. Curves I to IV, V$_u$ and
VI$_u$ are the $u(r)$ functions for the corresponding rays, curves V$_w$ and
VI$_w$ are the $w(r) = 1/u(r)$ functions. Such a change of variables was
necessitated by numerical problems, as mentioned above. The $w(r)$ curve for ray
VII is barely visible between I and II from $r = 0$ to nearly $r = 0.5$, then it
turns down and nearly coincides with curve VI$_w$. The corresponding $u(r)$
curve lies far above the upper margin of the figure. Rays 1b and VIII proceed
along the line $u = 0$. The rays that nearly coincide in Fig. \ref{axsymrays}
are widely separated in the $u$ direction.

The rays that have small $u_o$ stay close to the $u = 0$ line except near the
end point. This has consequences for the redshift profile; see below.

\begin{figure}[h]
\hspace{-5mm} \includegraphics[scale=0.64]{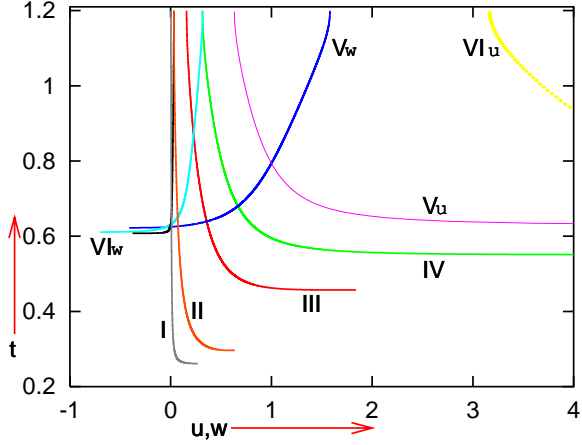} \caption{The $t(u)$
and $t(w)$ graphs for rays I -- VII. See text for more explanation.}
\label{utgraphs}
\end{figure}

Figure \ref{utgraphs} shows the $t(u)$ and $t(w)$ graphs for rays I -- VII. The
labels follow the same rules as in Fig. \ref{urgraphs}. All rays have their
upper ends at the same $t = 1.2$ because this is the $t$-coordinate of their
initial points. As before, the $t(w)$ curve for ray VII is barely visible
between I and II from $t = 1.2$ down to $t \approx 0.6$, then it turns left and
nearly coincides with curve VI$_w$. The corresponding $t(u)$ curve lies far
beyond the right margin of the figure. Rays 1b and VIII proceed along the line
$u = 0$.

\begin{figure}[h]
\hspace{-8mm} \includegraphics[scale=0.64]{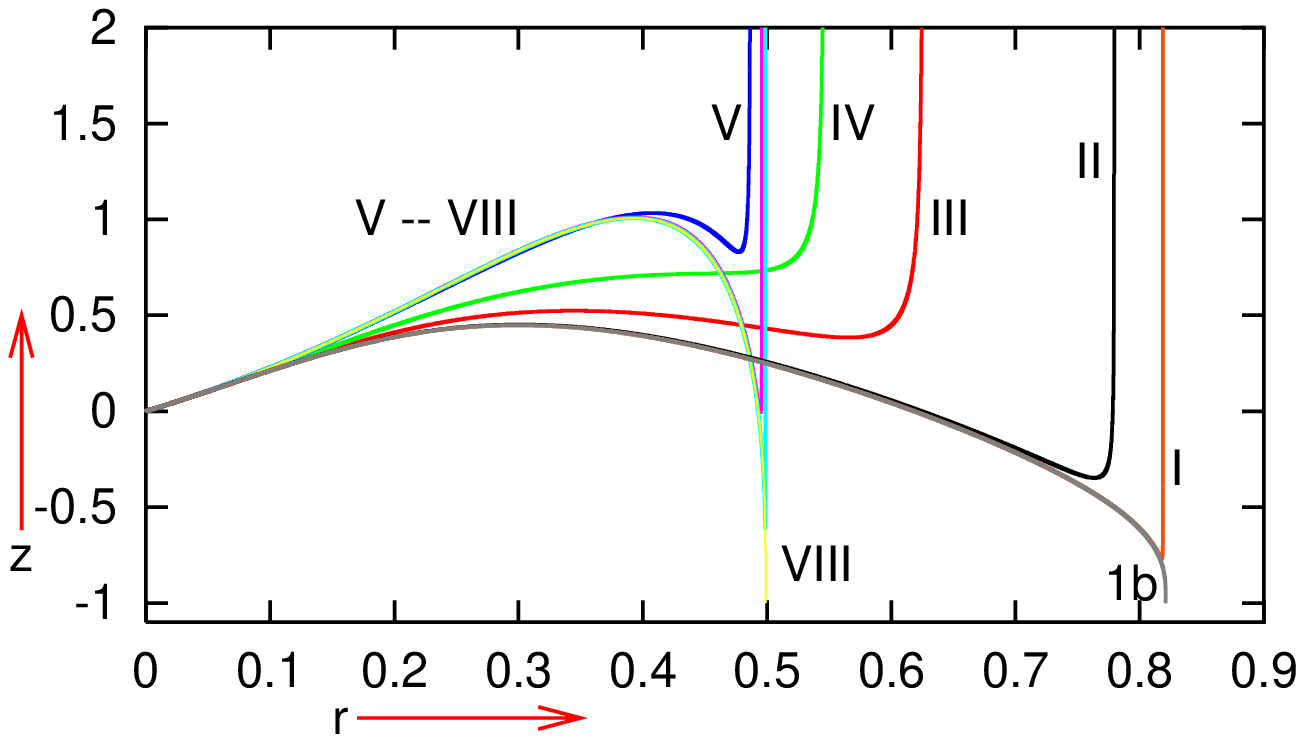}
\includegraphics[scale=0.5]{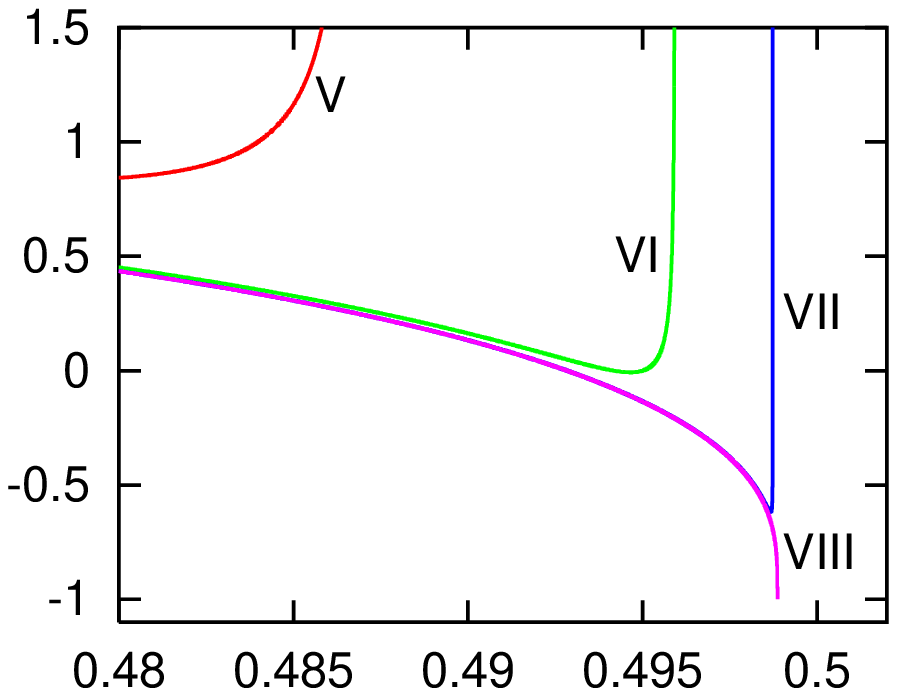}
\caption{{\bf Upper panel:} The $z(r)$ redshift profiles for rays I -- VIII. See
text for more explanation. {\bf Lower panel:} Enlarged view of profiles V to
VIII in the $r$-range where they are close together in the upper panel.}
\label{redsgraphs}
\end{figure}

Figure \ref{redsgraphs} shows the $z(r)$ profiles for all the rays. Only rays 1b
and VIII have $z \approx -1$ at the BB (see Table \ref{negz} for the actual
numerical values). Along the other rays $z$ first achieves a maximum, then
decreases, goes through a minimum and becomes very large near the BB. Table
\ref{negz} shows the minima of $1 + z$ along the rays.

\begin{table}[h]
\caption{Local minima of $1 + z$ on the rays from Fig. \ref{redsgraphs}}
\bigskip
\begin{tabular}{|l|l|}
  \hline \hline
\ \ Ray\ \ & \ \ Minimum $1 + z$ \ \ \\
  \hline \hline
 \ \ 1b\ & \ \ $2.53722758372994317 \times 10^{-6}$\ \\
  \hline
 \ \ I\ & \ \ 0.23305708892873711\ \\
  \hline
 \ \ II\ & \ \ 0.65207915516116299\ \\
\hline
 \ \ III\ & \ \ 1.3841852900906961\ \\
  \hline
 \ \ IV\ & \ \ 1\ \\
 \hline
 \ \ V\ & \ \ 1.8308297303885050\ \\
 \hline
 \ \ VI\ & \ \ 0.99266554207399438\ \\
 \hline
  \ \ VII\ & \ \ 0.38155727836044029\ \\
 \hline
  \ \ VIII & \ \ $8.27466387436995242 \times 10^{-6}$\ \\
  \hline \hline
\end{tabular}
\label{negz}
\end{table}

On rays I and VII, on which $u_o$ is near its``axial'' values $u = 0$ and $u =
\infty$, respectively, the minimum $z$ is near to $-1$, and the graphs suggest
that it is becoming still nearer to $z = -1$ when $u$ approaches zero or
infinity. This implies that on rays that go through the origin $r = 0$ with a
sufficiently small (or large) value of $u$, the point of minimum $z$ may lie
closer to the BB than the last scattering hypersurface (the Szekeres model does
not apply before last scattering because of $p = 0$). Such a ray would thus
display a finite blueshift to the present observer even if it were emitted
during last scattering.

The meaning of the initial $u$ on a ray going off from the origin becomes
clearer when $u$ is transformed, in a surface of constant $r$, by
\begin{equation}\label{11.16}
u = S \tan (\vartheta/2).
\end{equation}
Then the metric of a surface of constant $t$ and $r$ in (\ref{2.1}) becomes
${\rm d} {s_2}^2 = \Phi^2 \left({\rm d} \vartheta^2 + \sin^2 \vartheta {\rm d}
\varphi^2\right)$. The value $u = 0$ corresponds to $\vartheta = 0$, the limit
$u \to \infty$ corresponds to $\vartheta \to \pi$. Thus, the only null geodesics
that can display infinite blueshifts to an observer at $r = 0$ are those that
reach her tangentially to the directions $\vartheta = 0$ and $\vartheta = \pi$.
Geodesics approaching this observer from any other direction will display finite
blueshifts or, if emitted at the BB, infinite redshifts.

\begin{figure}[h]
\hspace{-5mm} \includegraphics[scale=0.64]{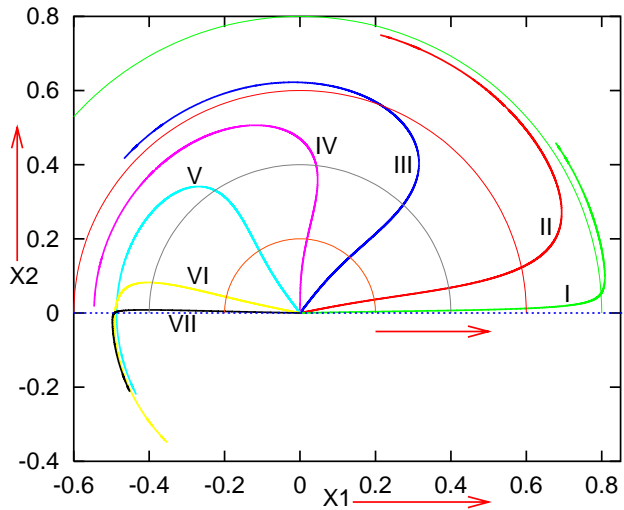} \caption{Rays I --
VIII projected along the $r =$ constant lines on the surface of constant $t$ and
$\varphi$ in the polar coordinates (\ref{11.16}). See explanations in the text.}
\label{polar}
\end{figure}

Figure \ref{polar} shows the projections of rays I -- VIII on the surface of
constant $t$ and $\varphi$ along the $r =$ constant lines. The coordinates of
the graph are $X_1 = r \cos \vartheta$ and $X_2 = r \sin \vartheta$. The arrow
within the figure is parallel to the half-line $u = \vartheta = 0$, on which the
dipole component of mass density has maxima. The dotted line $X_2 = 0$ is the
projection of the axial rays 1b (going from $(X_1, X_2) = (0, 0)$ to the right)
and VIII (going from $(0, 0)$ to the left). The thin circles are curves of
constant $r$.

For clarity, Fig. \ref{polar} does not show the mirror-reflections of the rays
in the $X_2 = 0$ axis. One obtains the collection of rays running in all the
$\varphi =$ constant hypersurfaces by rotating the plane of Fig. \ref{polar}
around the $X_2 = 0$ axis. Thus, rays going off from $r = 0$ with initial $u >
0$ (i.e. $\vartheta > 0$) and $u < \infty$ ($\vartheta < \pi$) form a funnel
around the direction $u = 0$; only the rays with $u = 0$ and $u = \infty$ remain
axial all the way and display infinite blueshifts to the observer when they
reach the BB.

Note that rays running close to the half-line $\vartheta = 0$ (i.e. $u = 0$)
bend away from it when approaching the BB, while those running close to the
half-line $\vartheta = \pi$ ($u = \infty$) bend toward it at the BB. We will see
this pattern repeated in the nonsymmetric model in Sec. \ref{simgen}.

Rays II -- VII behave similarly to nonradial rays in an L--T model: when they
approach the BB where $\dril {t_B} r \neq 0$, they bend sideways and hit the BB
tangentially to a hypersurface of constant $r$ (see Fig. 10 in Ref.
\cite{Kras2016}). But in L--T models any ray passing through the center $r = 0$
is radial, so all the rays shown in this figure would become radial in the L--T
limit. Thus, the limiting transition from a Szekeres to an L--T model is
discontinuous, just like the transition from L--T to Friedmann, in which all
blueshifts discontinuously disappear.

Figures \ref{redsgraphs} and \ref{polar} show that the strongly blueshifted (SB)
rays are centers of instability. Consider a single SB ray. The rays that are
near to it develop negative $z$, but before they hit the BB, $z$ goes through a
minimum $> -1$, and then increases to infinity. As the neighbouring rays move
nearer to the SB ray, the minimum becomes closer to $-1$, but $z$ at the BB is
still infinite, and the ray still ends up being tangent to the BB and to an $r =
$ constant contour. Rays on opposite sides of the SB ray bend in opposite
directions. Only the SB ray has $z = -1$ at the BB, and hits the BB orthogonally
to an $r = $ constant contour. In the axisymmetric case we knew which rays would
be SB, so we forced the numerical program to keep the ray on the SB path
exactly. However, in the general case we do not know where the SB paths are, and
this instability will render finding the SB paths numerically impossible: we can
only approach them and observe the characteristic features described above; see
Sec. \ref{simgen}.

In Fig. \ref{polar} the SB rays are orthogonal to the contours of constant $r$
on approach to the BB, and the projections of the SB rays on the constant-$(t,
\varphi)$ hypersurface are tangent to the dipole axes at the contact with the
BB. (They are in fact tangent to these axes all the way.) We will see in Sec.
\ref{simgen} that the first property is nearly reproduced in the nonsymmetric
QSS model, while the second one does not survive.

\section{Null geodesics in a simple example of a general QSS
model}\label{simgen}

\setcounter{equation}{0}

We shall now consider null geodesics going off from the origin to the past in a
simple QSS model that has no symmetry. In order to keep calculations simple, and
to stay close to the axially symmetric example of Sec. \ref{exQSS}, we still
assume $M$, $E$, $S$ and $t_B$ of the forms (\ref{3.2}), (\ref{8.5}),
(\ref{9.2}) and (\ref{9.4}), respectively. We choose $P(r)$ and $Q(r)$ so that
(\ref{2.6}) and (\ref{2.7}) are fulfilled in the simplest way possible. These
two inequalities, with (\ref{3.2}), (\ref{8.5}) and (\ref{9.2}) are equivalent
to the following one:
\begin{equation}\label{12.1}
\left(P,_r\right)^2 + \left(Q,_r\right)^2 < \frac {a^2} {r^2} + \frac {a^2} {a^2
+ r^2}.
\end{equation}
In order to avoid any symmetries, $(P, Q, S)$ must be linearly independent. We
choose
\begin{equation}\label{12.2}
P(r) = \frac {pa} {2 \left(a^2 + r^2\right)}, \qquad Q(r) = \frac {qa}
{\sqrt{a^2 + r^2}},
\end{equation}
where $p$ and $q$ are constant parameters. The functions $(P, Q, S)$ were chosen
such that $P,_r(0) = Q,_r(0) = S,_r(0) = 0$. This is needed to cancel the
infinity in ${\cal N}/\Phi$ at $r = 0$ in (\ref{5.5}) and (\ref{5.6}) by $({\cal
E},_r/{\cal E}),_x$ and $({\cal E},_r/{\cal E}),_y$. Equation (\ref{12.1}) will
be obeyed at all $r$ if
\begin{equation}\label{12.3}
p^2 < 4a^4, \qquad q^2 < 5a^2;
\end{equation}
see Appendix \ref{condpq} for a proof. Since in most of the examples so far we
had $a^2 = 0.1$, we now choose
\begin{equation}\label{12.4}
(a^2, p, q) = (0.1, 0.15, 0.6).
\end{equation}

At any chosen $r$, the extrema of the dipole component of the mass-density (see
comment under (\ref{2.8})) occur where $\left({\cal E},_r/{\cal E}\right),_x =
0$ and $\left({\cal E},_r/{\cal E}\right),_y = 0$, i.e. where
\begin{eqnarray}\label{12.5}
&& x - P = - SP,_r / {\cal W}, \qquad y - Q = - SQ,_r / {\cal W}, \nonumber \\
&& {\cal W} \df S,_r + \varepsilon \sqrt{\left(P,_r\right)^2 +
\left(Q,_r\right)^2 + \left(S,_r\right)^2}, \nonumber \\ && \varepsilon \df \pm
1.
\end{eqnarray}
The value of ${\cal E},_r/{\cal E}$ at these extrema is
\begin{equation}\label{12.6}
\left({\cal E},_r/{\cal E}\right)_{\rm ex} = \varepsilon
\sqrt{\left(P,_r\right)^2 + \left(Q,_r\right)^2 + \left(S,_r\right)^2} / S,
\end{equation}
and so $\varepsilon = +1$ corresponds to the maximum of ${\cal E},_r/{\cal E}$,
while $\varepsilon = -1$ corresponds to the minimum.

\begin{figure}[h]
\includegraphics[scale=0.8]{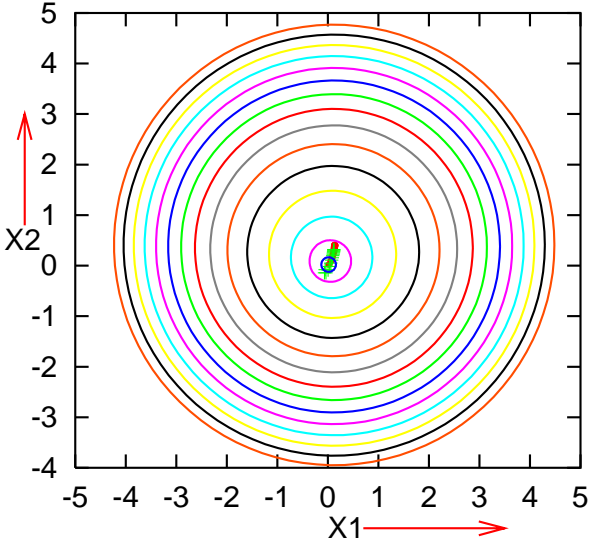}
\caption{Spheres of constant $(t, r)$ in the nonsymmetric Szekeres model
(\ref{2.1}) -- (\ref{2.3}) with $P$, $Q$ and $S$ given by (\ref{12.2}),
(\ref{12.4}) and (\ref{9.2}). This is a projection on the $(X_1, X_2)$ Cartesian
plane. See explanation in text.} \label{xyproj}
\end{figure}

\begin{figure}[h]
\includegraphics[scale=0.8]{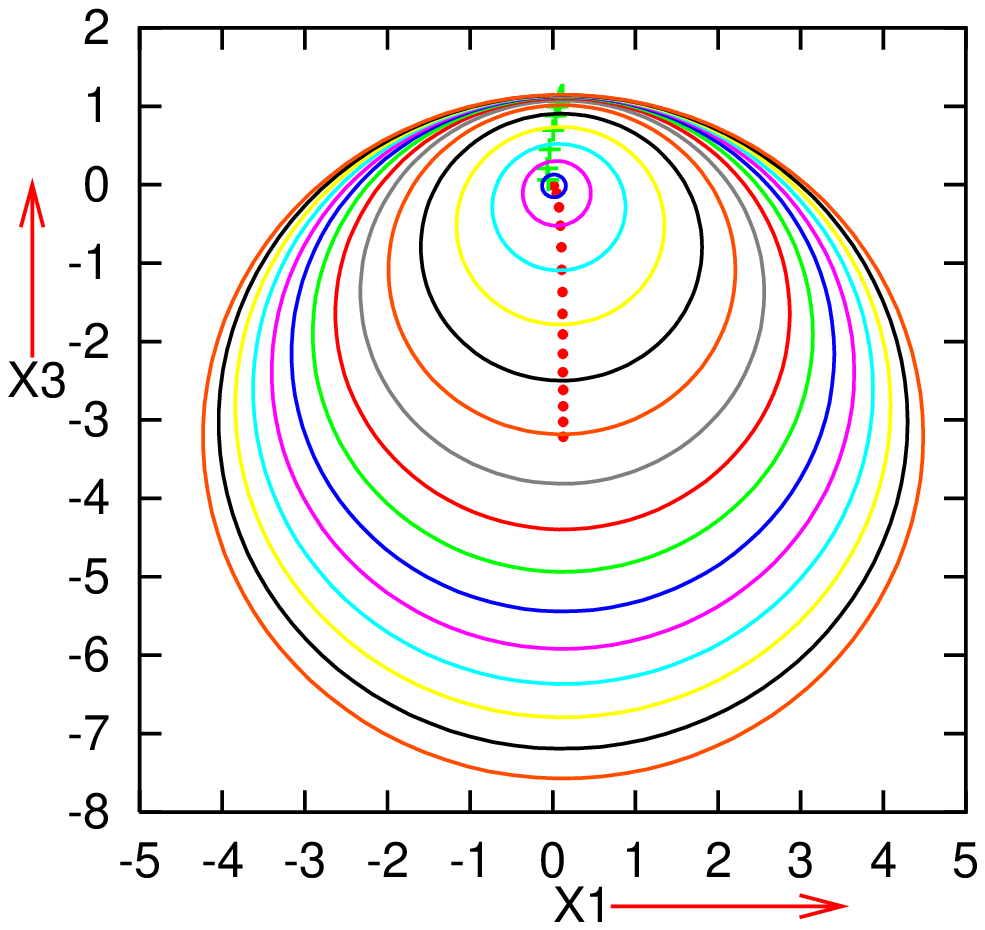}
\caption{The same spheres as in Fig. \ref{xyproj} projected on the $(X_1, X_3)$
Cartesian plane.} \label{xzproj}
\end{figure}

\begin{figure}[h]
\includegraphics[scale=0.8]{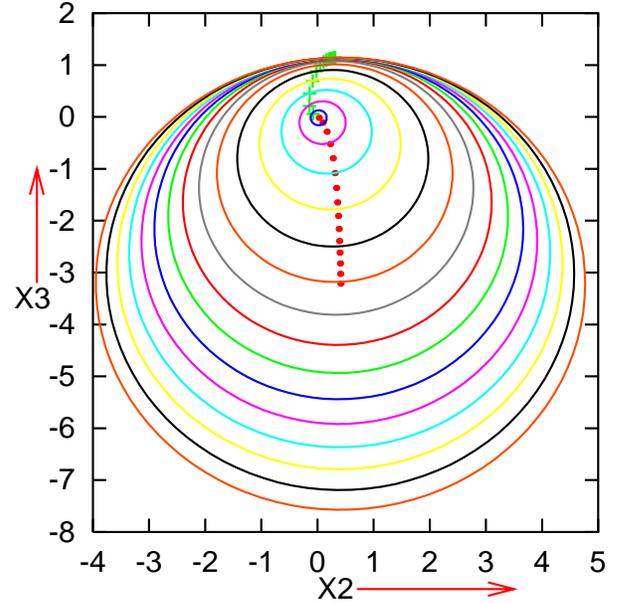}
\caption{The same spheres as in Fig. \ref{xyproj} projected on the $(X_2, X_3)$
Cartesian plane.} \label{yzproj}
\end{figure}

The positions of the $r = $ constant spheres in a $t =$ constant space are
illustrated in Figs. \ref{xyproj} -- \ref{yzproj}. The values of the
$r$-coordinate on the spheres range from $r = 0.2$ to $r = 3$ at intervals of
$\Delta r = 0.2$. The five largest spheres are in the Friedmann region. The
family of spheres was first mapped into an Euclidean space in such a way that
the shortest Euclidean distances between them are the same as the shortest
geodesic distances in the Szekeres spacetime; see Sec. \ref{exQSS}. These
figures represent projections of the spheres on the $(X_1, X_2)$, $(X_1, X_3)$
and $(X_2, X_3)$ Euclidean coordinate planes. The centers of the spheres are
marked with dots, the positions of the dipole maxima on the spheres are marked
with crosses. The crosses do lie on the spheres, but not on their outer edges
seen from the three directions. This is why they project into the interiors of
the great circles, which is most conspicuous in Fig. \ref{xyproj}. See Appendix
\ref{projcalc} for information on the calculations underlying Figs. \ref{xyproj}
-- \ref{yzproj}.

Using (\ref{4.11}) we find
\begin{equation}\label{12.7}
\lim_{r \to 0} \left(\frac {P,_r} {\Phi}\right) = - \frac p {a^3 R}, \qquad
\lim_{r \to 0} \left(\frac {Q,_r} {\Phi}\right) = - \frac q {a^2 R}.
\end{equation}

{}From (\ref{12.7}) and (\ref{11.1}) it follows that the coefficients of
$(k^r)^2$ in (\ref{5.5}) and (\ref{5.6}) have finite limits at $r \to 0$, so the
ones to take care about are the terms containing $k^r k^x$ and $k^r k^y$.
Equation (\ref{5.7}) implies that $k^r \neq 0$ at $r = 0$ because otherwise also
$k^t = 0$ at $r = 0$, and the geodesic would be spacelike at this point. This
means that the terms in question can be finite at the origin only if $k^x = k^y
= 0$ there. Thus we can apply the de l'H\^opital rule to calculate the limits at
$r \to 0$ of $k^x/\Phi$ and $k^y/\Phi$. We do it in the same way as in
(\ref{11.3}): we calculate these limits along the ray, and in the second step
use (\ref{5.5}) and (\ref{5.6}) to substitute for $\dril {(k^x)} {\lambda}$ and
$\dril {(k^y)} {\lambda}$. The results are, using (\ref{4.9}) -- (\ref{4.11}):
\begin{eqnarray}
&& \lim_{r \to 0} \frac {k^x} {\Phi} = \frac 1 {3R} \lim_{r \to 0} \frac {{\cal
E},_r {\cal E},_x - {\cal E} {\cal E},_{rx}} {\Phi} \nonumber \\
&& = \frac 1 {3a^2 R^2} \left\{x - \frac {3p} {4a} + \frac q {a^2}\ \left(x -
\frac p {2a}\right) (y - q)\right. \nonumber \\
&& \ \ \ + \left.\frac p {4 a^3}\ \left[\left(x - \frac p {2a}\right)^2 - (y -
q)^2\right]\right\},\ \ \ \ \label{12.8} \\
&& \lim_{r \to 0} \frac {k^y} {\Phi} = \frac 1 {3R} \lim_{r \to 0} \frac {{\cal
E},_r {\cal E},_y - {\cal E} {\cal E},_{ry}} {\Phi} \nonumber \\
&& = \frac 1 {3a^2 R^2} \left\{y - \frac {3q} 2 + \frac p {2 a^3} \left(x -
\frac p {2a}\right) (y - q)\right. \nonumber \\
&& \ \ \ + \left.\frac q {2 a^2} \left[- \left(x - \frac p {2a}\right)^2 + (y -
q)^2\right]\right\}.\ \ \ \ \label{12.9}
\end{eqnarray}

\begin{figure}[h]
\hspace{-5mm} \includegraphics[scale=0.7]{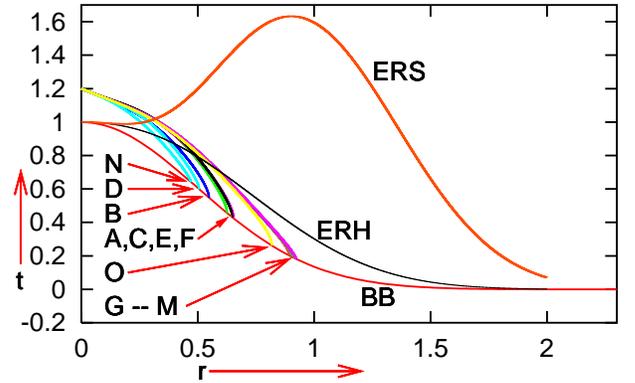}
\caption{The $t(r)$ graphs of rays A -- O.}
\label{nosymcones}
\end{figure}

\begin{figure}[h]
\includegraphics[scale=0.64]{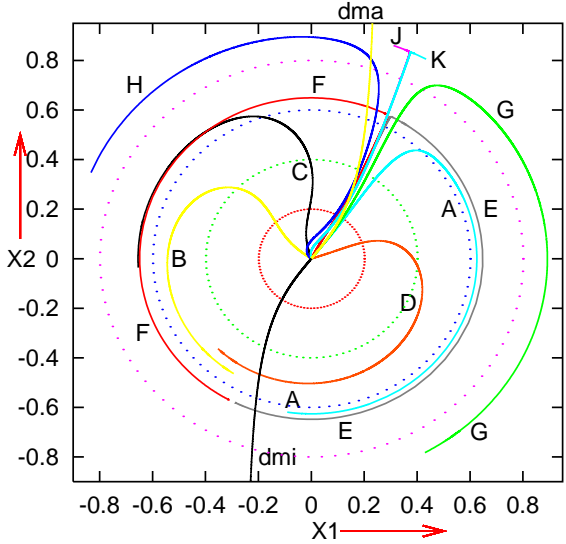}
\caption{The $r(\varphi)$ graphs of rays A -- K (in polar coordinates). See text
for explanations.} \label{polphi1}
\end{figure}

\begin{figure}[h]
\includegraphics[scale=0.5]{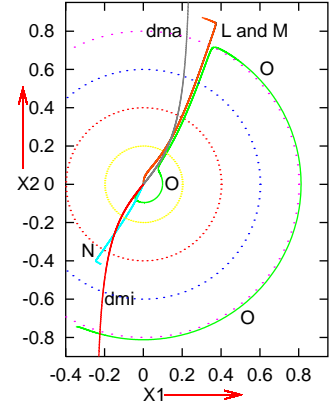}
\caption{The $r(\varphi)$ graphs of rays L -- O (in polar coordinates). Rays L
and M nearly coincide, except that L goes farther. See text for explanations.}
\label{polphi2}
\end{figure}

\begin{figure}[h]
\includegraphics[scale=0.64]{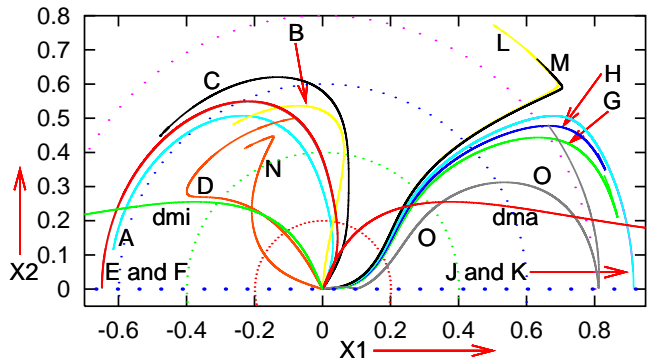}
\caption{The $r(\vartheta)$ graphs of rays A -- O.}
\label{polthe}
\end{figure}

\begin{figure}
\hspace{-7mm} \includegraphics[scale=0.61]{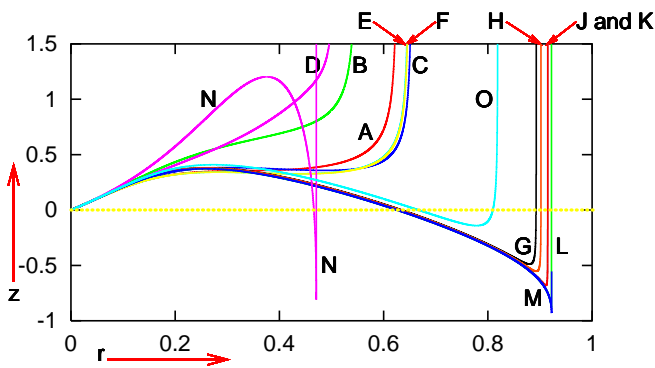} \caption{The $z(r)$
redshift profiles for rays A -- O. The dotted horizontal line marks $z = 0$.}
\label{nosymz}
\end{figure}

Figures \ref{nosymcones} -- \ref{nosymz} show the results of integrating
(\ref{5.1}) -- (\ref{5.7}) with $P$, $Q$ and $S$ given by (\ref{12.2}) and
(\ref{9.2}), and the values of the parameters given by (\ref{12.4}) and
(\ref{10.1}). The numerical procedure was the following:

1. The values of $x$ and $y$ at the maximum and at the minimum of the dipole
were calculated from (\ref{12.5}) for every $r$ in the range $[0, 2]$ with the
step $\Delta r = 1/300\ 000$.

2. The initial values of $r$ and $t$ were the same as for ray 1b in Sec.
\ref{simpleblue}:
\begin{equation}\label{12.10}
(r, t)_o = (0, 1.2).
\end{equation}

3. The initial values of $x$ and $y$ on the rays were chosen by trial and error,
so as to approach the direction in which strong blueshifts can be expected.

4. As explained in the paragraph below (\ref{12.7}), the initial values of $k^x$
and $k^y$ were chosen zero.

5. For each pair $(x_o, y_o)$, the initial values of $(k^x/\Phi)_o$ and
$(k^x/\Phi)_o$ were calculated from (\ref{12.8}) and (\ref{12.9}).

6. The initial value of $k^t$ was taken $-1$, in agreement with (\ref{5.8}). The
initial value of $k^r$ was then calculated from (\ref{5.7}), knowing that $k^x_o
= k^y_o = 0$ and $k^r > 0$ (at the center $r = 0$ there is no other possibility
for $k^r$).

7. The step in the affine parameter was chosen $\Delta \lambda = 10^{-7}$. The
initial value of $\lambda$ was irrelevant, since $\lambda$ does not appear
explicitly in any of the graphs.

8. Given the above, the next value of each function $f(\lambda)$ was calculated from
\begin{equation}\label{12.11}
f(\lambda + \Delta \lambda) = f(\lambda) + \dr f {\lambda}\ \Delta \lambda.
\end{equation}

9. At each $\lambda$, $\dril {(t, r, x, y)} {\lambda}$ were calculated from
(\ref{5.1}), $\dril {(k^t, k^x, k^y)} {\lambda}$ were calculated from
(\ref{5.3}) and (\ref{5.5}) - (\ref{5.6}). With these data, the values of $(t,
r, x, y, k^t, k^x, k^y)$ at $\lambda + \Delta \lambda$ were found using
(\ref{12.11}), and the value of $k^r$ at $\lambda + \Delta \lambda$ was found
from (\ref{5.7}) assuming $k^r > 0$.

The values of $x$ and $y$ at the dipole maximum and minimum at $r = 0$ are
\begin{eqnarray}\label{12.12}
&& x^o_{\rm max} = 0.36830403011403989, \nonumber \\
&& y^o_{\rm max} = 0.76587184263184405, \nonumber \\
&& x^o_{\rm min} = -5.61304662256229547 \times 10^{-2}, \nonumber \\
&& y^o_{\rm min} = 0.22899995223995073.
\end{eqnarray}
Table \ref{table3} gives the initial values of $x$ and $y$ for the various rays
shown in the figures and the local minimum of $1 + z$ achieved on each ray.
Where the smallest $1 + z = 1$, the redshift along the ray was monotonically
increasing.

\begin{table*}
\caption{Properties of the rays from Figs. \ref{nosymcones} -- \ref{nosymz}.}
\bigskip
\begin{tabular}{|l|l|l|l|}
  \hline \hline
\ Ray\ & \ Initial $x$ \ & \ Initial $y$ \ & \ min. $1 + z$ \\
  \hline \hline
 \ A & \ $x^o_{\rm max}$ & \ $y^o_{\rm max}$ & \ 1\ \\
  \hline
 \ B & \ $0$ & \ $y^o_{\rm max}$ & \ 1\ \\
  \hline
 \ C & \ $x^o_{\rm max}/2$ & \ $y^o_{\rm max}$ & \ 1.3576\ \\
\hline
 \ D & \ $2 x^o_{\rm max}$ & \ $y^o_{\rm max}$ & \ 1\ \\
  \hline
 \ E & \ $0.88 \times x^o_{\rm max}$ & \ $y^o_{\rm max}$ & \ 1.3274\ \\
 \hline
 \ F \ & \ $0.875 \times x^o_{\rm max}$ & \ $y^o_{\rm max}$ & \ 1.3261\ \\
 \hline
 \ G & \ $p/(2a)$ & \ $q$ & \ 0.511417\ \\
 \hline
  \ H & \ $0.47p/a + 0.012 x^o_{\rm max}$\ & \ $0.94q + 0.06 y^o_{\rm max}$ &
  \ 0.44733274\ \\
 \hline
  \ J & \ $0.47p/a + 0.03312 x^o_{\rm max}$\ & \ $0.94q + 0.06 y^o_{\rm max}$ &
  \ 0.319953\ \\
 \hline
  \ K &\ $0.47p/a + 0.03324 x^o_{\rm max}$\ & \ $0.94q + 0.06 y^o_{\rm max}$ &
  \ 0.3194256\ \\
 \hline
  \ L &\ $0.405 p/a + 0.0585 x^o_{\rm max}$\ & \ $0.81q + 0.09 y^o_{\rm max}$ &
  \ 0.12313\ \\
 \hline
  \ M &\ $0.456 p/a + 0.0567776 x^o_{\rm max}$\ & \ $0.912q + 0.088 y^o_{\rm
  max}$ & \ 0.06946378\ \\
 \hline
  \ N &\ $-2 p/a - 6.51 x^o_{\rm min}$\ & \ $-4 q + 5 y^o_{\rm min}$ & \
  0.19046\ \\
 \hline
  \ O &\ $-0.47 p/a - 0.0324 x^o_{\rm min}$\ & \ $0.94 q + 0.06 y^o_{\rm min}$
  & \ 0.8589\ \\
  \hline \hline
\end{tabular}
\label{table3}
\end{table*}

The initial points of the rays come in 4 sets. The first set consists of rays A
-- F, which had initial $y$ at $y^o_{\rm max}$, and initial $x$ varied by trial
and error around $x^o_{\rm max}$.

The second set consists of rays G -- K, for which the initially chosen $x_1 -
p/(2a)$ and $y_1 - q$ were multiplied by $\alpha$ to obtain
\begin{equation}\label{12.13}
[x_2 - p/(2a), y_2 - q] = \alpha [x_1 - p/(2a), y_1 - q].
\end{equation}
This was meant to change $\sqrt{[x - p/(2a)]^2 + (y - q)^2}$ while keeping $(y -
q)/[x - p/(2a)]$ unchanged (this meant changing $\vartheta$ while keeping
$\varphi$ unchanged in the polar graphs in Figs. \ref{polphi1}, \ref{polphi2}
and \ref{polthe}). The variable parameters were $x_1$ and $\alpha$, both chosen
by trial and error. For ray G $\alpha = 0$, for the other rays in this group
$\alpha = 0.06$. The parameter $x_1$ on rays H -- K was $\beta x^o_{\rm max}$,
with $\beta$ being 0.2 on ray H, 0.552 on ray J and 0.554 on ray K.

The third set consists of rays L and M. They have minimum $z$ closer to $-1$
than on the other rays. For ray L $\alpha = 0.09$ and $\beta = 0.65$; for ray M
$\alpha = 0.088$ and $\beta = 0.6452$. Attempts at nailing down the expected SB
ray with larger precision were becoming prohibitively time-consuming and were
abandoned.

Rays A -- M had their initial directions in a vicinity of the dipole maximum.
The rays in the fourth set, N and O, had their initial directions in a vicinity
of the dipole minimum. For ray N, $\alpha = 5$ and $x_1 = -1.33 x^o_{\rm min}$;
for ray O $\alpha = 0.06$ and $x_1 = 0.554 x^o_{\rm min}$. Ray N was meant to
demonstrate that a $z(r)$ profile indicating proximity to an SB ray exists also
in a vicinity of the dipole minima. Attempts at further improvement were
abandoned when they became too time-consuming.

Ray O is an example of a weird behaviour caused by proximity to an SB ray; see
further below.

Figure \ref{nosymcones} shows the $t(r)$ graphs of rays A -- O. The ERS was
calculated only for axial rays in the axially symmetric case, so it is an
extraneous element in the general case. But it corresponds to the same BB
profile, so is drawn for comparison. The ERH is the analogue of the ERS in the
L--T limit, also calculated for the same BB profile.

Figures \ref{polphi1} and \ref{polphi2} show the $r(\varphi)$ graphs of all the
rays, where $\varphi(r)$ and $\vartheta(r)$ are related to $x(r)$ and $y(r)$ by
(\ref{2.3}). The coordinates in these figures are $X_1 = r \cos \varphi$ and
$X_2 = r \sin \varphi$, where $\tan \varphi = (y - Q)/(x - P)$. These are
abstract graphs that are not projections of the rays on any actual subspace of
the Szekeres manifold. They illustrate the process of approaching the expected
SB paths.

The dotted circles are curves of constant $r$. The lines marked 'dma' and 'dmi'
are the positions of the dipole maximum and minimum, respectively, for each $r$.
Figure \ref{polphi1} shows the paths of the rays that have their initial points
in the vicinity of 'dma' at $r = 0$; call them rays of class 1. Curve A had the
initial direction tangent to the 'dma' path and was the first trial. Rays B -- K
show consecutive approximations to the SB path obtained by moving the initial
$x$ and $y$ by trial and error around $x^o_{\rm max}$ and $y^o_{\rm max}$. The
SB path should lie near to J and K.

Figure \ref{polphi2} shows four more rays; they are drawn separately to avoid
clogging the image. Curves L and M are the paths of class 1 rays that hit the BB
at still larger $r$ than J and K, and are still better approximations to the
expected SB path, with curve M being the best approximation; see Table
\ref{table3} and Fig. \ref{nosymz} further below. Curves N and O have their
initial $(x, y)$ in the vicinity of 'dmi' at $r = 0$, call them rays of class 2.
Curve N is a good approximation to a second SB path.

The instability around the SB ray in class 2 is different than in class 1. In
class 1, rays close to the SB path bend away from it, while in class 2 they bend
toward it. We observed the same characteristic pattern in Fig. \ref{polar}.

Curve O is an example of a strange path that results when the initial $(x, y)$
in class 2 is close to $[p/(2a), q]$ and the initial direction is close to the
'dmi' path. This curve begins close to 'dmi', then bends away from it, goes
around half a circle at nearly constant $r$, then proceeds near the opposite SB
path, and finally, near the BB, bends again to once more go around half a circle
at nearly constant $r$. The minimum redshift along it is moderately negative;
see Table \ref{table3} and Fig. \ref{nosymz}.

Rays M and N which are expected to be near to the SB rays do meet the BB nearly
orthogonally to the $r =$ constant surfaces, but are not related to the dipole
maxima and minima, unlike in the axially symmetric case in Fig. \ref{polar}.
Thus, the coincidence between the SB rays and the dipole extrema in that case
was forced by the symmetry of the model, and is not generic. This is still one
more warning that by studying models with symmetries we may have forced
independent objects to coincide.\footnote{Well-known examples of this phenomenon
are the stationary limit hypersurface in the Kerr metric \cite{Cart1973} and the
apparent horizon in the L--T models \cite{KrHe2004}, which are in general
distinct from the event horizon, but coincide with it in the Schwarzschild
limit.\label{memfoot}}

It follows that the SB rays determine preferred directions in general
quasi-spherical Szekeres spacetimes that are independent of the trajectories of
the dipole extrema.

Figure \ref{polthe} shows the $r(\vartheta)$ graphs of all the rays. Like Figs.
\ref{polphi1} and \ref{polphi2}, this is an abstract graph which is not supposed
to represent a projection of the rays on any actual subspace. The horizontal
dotted line is $\vartheta = 0$. By definition, $\vartheta$ must obey $0 \leq
\vartheta \leq \pi$, so no curve can cross the line $\vartheta = 0$, and this is
why curve O bounces off it close to the point $r = 0$. The sharp turns on curves
G, H, L, M, N and O occur where they are near the BB and begin to approach it
tangentially to a surface of constant $r$. In this graph, the curves that are
supposed to be near to the SB paths (L, M and N) do not approach the circular
segment orthogonally to the $r = $ constant lines.

Finally, Fig. \ref{nosymz} shows the $z(r)$ profiles along all the rays. The
profiles along rays G -- M form a sequence similar to \{II, I\} in Fig.
\ref{redsgraphs} that approached the 1b profile. At the scale of Fig.
\ref{nosymz} M overlays L, but, as seen from Table \ref{table3}, has a smaller
minimum of redshift. The profile along ray N is similar to profile VIII; both 1b
and VIII were shown to be the actual strongly blueshifted rays. Consequently,
rays M and N must be near to SB rays.

\section{Conclusions and summary}\label{sumup}

\setcounter{equation}{0}

After preliminary discussions and definitions in Secs. \ref{QSSS} -- \ref{ERS},
we introduced an exemplary axially symmetric quasi-spherical Szekeres (QSS)
spacetime in Sec. \ref{exQSS}. We then showed numerically in Sec.
\ref{simpleblue} that light rays exist in it that display strong blueshifts ($1
+ z \approx 10^{-6}$) when they are emitted near the Big Bang (BB). The strongly
blueshifted (SB) rays intersect every space of constant $t$ on the axis of
symmetry. In this case, the axes of the mass dipoles coincide with the axis of
symmetry -- and, automatically, also with the spatial paths of the SB rays.

In an L--T model, if a hump in the BB generates blueshift, then the blueshift is
the same along rays passing through the center in all directions. In the axially
symmetric Szekeres model, as the numerical computations show, the blueshift
would be strong only along such rays whose directions at origin are tangent or
near to the surfaces $u = 0$ and $u = +\infty$. Consequently, a present observer
could see the blueshift only if she were placed near or in one of these
surfaces.

The collection of rays emitted at the BB in an L--T model, on which the observed
$z = -1$, is labelled by three parameters: two of them (the initial values of
$x$ and $y$) define the radial direction, the third one defines the radial
coordinate $r_e$ of the emission point on the BB where $t_e = t_B(r_e)$. In the
axially symmetric model (\ref{7.4}) -- (\ref{7.5}), as shown by the numerical
experiments in Sec. \ref{simpleblue} and \ref{moreblue}, the collection of such
rays is labelled by two parameters. One parameter is the value of $\varphi$
(which is constant on each ray), the other parameter, the initial value of $r$,
determines the emission point of the ray at the BB. So, the collection of all
rays that have strong blueshifts is in this case a two-dimensional set.

Each SB ray is unstable: rays in every small neighbourhood of it all have
infinite redshift between the BB and any later observer. Therefore, tracking the
SB rays numerically is a rather hopeless undertaking: in the axially symmetric
case we knew in advance where to expect them, and the condition of staying on
the SB path was built into the numerical code. Nevertheless, such an attempt was
undertaken in Sec. \ref{simgen}, in a QSS model without any symmetry. It was
shown that sequences of rays exist in it, along which the redshift profiles are
similar to those on rays that approach an SB ray in the axially symmetric case.
The tentative conclusion is that SB rays should exist also in general QSS
models. The numerical calculations show clearly that in the nonsymmetric case
the spatial paths of the SB rays are not related to the paths of the mass dipole
extrema. Thus, the SB rays define other preferred directions in general QSS
spacetimes.

Preferred null directions immediately evoke the association with principal null
directions (PNDs) of the Weyl tensor. But in the nonsymmetric QSS model the SB
rays do not coincide with the PNDs. The Weyl tensor of a general Szekeres
spacetime is of Petrov type D, and in the coordinates of (\ref{2.1}) its PNDs
are
\begin{equation}\label{13.1}
k^{\alpha}_{\pm} = \left(1, \pm \frac {\sqrt{1 + 2E}} {\Phi,_r - \Phi {\cal
E},_r/{\cal E}}, 0, 0\right).
\end{equation}
These are geodesic only in the axially symmetric case; then they coincide with
the $\{u = 0, k^{\varphi} = 0\}$ rays of Sec. \ref{simpleblue} (they remain
geodesic and SB also in the L--T limit). Consequently, in a general model the SB
rays are independent of the PNDs (this is one more example to the phenomenon
mentioned in footnote \ref{memfoot}).

It remains a challenge to prove by exact methods that SB rays do exist in every
QSS spacetime. Incidentally, such a proof is still lacking even in the L--T
limit. It was proved exactly that rays emitted at the BB nonradially generate
infinite redshifts \cite{Kras2016}, so rays emitted at the BB {\it can possibly}
generate infinite blueshifts only if they are radial. But no exact proof exists
that indeed $1 + z = 0$ on those rays from the BB that do not prohibit it. All
that is available is a casual remark without proof by Szekeres \cite{Szek1980}
and confirmations of it by perturbative calculations \cite{HeLa1984}, and by
numerical calculations in selected explicit examples of L--T spacetimes
\cite{Kras2014,Kras2016}.

The toy models considered in this paper were used to demonstrate the following
three facts that should be useful in constructing realistic QSS models:

1. Rays with strong blueshifts do exist in an axially symmetric QSS model, and
most probably also exist in QSS models without any symmetry.

2. The blueshift generated in these models is strongly anisotropic (it exists
around only two directions).

3. The function $S$ in (\ref{2.2}) allows one to increase the size of the
blueshift-generating region without changing the BB profile.

\appendix

\section{Solutions of (\ref{2.4}) with $\Lambda = 0$}\label{freqs}

\setcounter{equation}{0}

With $\Lambda = 0$, the solutions of (\ref{2.4}) are the following:

when $E(r) < 0$:
\begin{eqnarray}\label{a.1}
\Phi(t,r) &=& - \frac M {2E} (1 - \cos \eta), \nonumber \\
\eta - \sin \eta &=& \frac {(-2E)^{3/2}} M \left[t - t_B(r)\right];
\end{eqnarray}
when $E(r) = 0$:
\begin{equation}\label{a.2}
\Phi(t,r) = \left\{\frac 9 2 M(r) \left[t - t_B(r)\right]^2\right\}^{1/3};
\end{equation}
when $E(r) > 0$:
\begin{eqnarray}\label{a.3}
\Phi(t,r) &=& \frac M {2E} (\cosh \eta - 1), \nonumber \\
\sinh \eta - \eta &=& \frac {(2E)^{3/2}} M \left[t - t_B(r)\right].
\end{eqnarray}
Throughout this paper only the case $E > 0$ is considered.

\section{The Riemann tensor for (\ref{2.1})}\label{riemann}

\setcounter{equation}{0}

In the orthonormal tetrad defined by the metric (\ref{2.1}), the tetrad
components of the Riemann tensor are:
\begin{eqnarray}
R_{0101} &=& - \frac M {\Phi^3} + \frac {3 M \Phi,_r} {\Phi^3 {\cal N}} - \frac
{M,_r} {\Phi^2 {\cal N}}, \\
R_{0202} &=& R_{0303} = \tfrac 1 2 R_{2323} = - \frac M {\Phi^3}, \\
R_{1212} &=& R_{1313} = - \frac {2 M} {\Phi^3} + \frac {3 M \Phi,_r} {\Phi^3
{\cal N}} - \frac {M,_r} {\Phi^2 {\cal N}},
\end{eqnarray}
where ${\cal N}$ is given by (\ref{5.2}). Note that at the origin, where
$\Phi/{\cal E} = 0$, if $|{\cal E},_r| < \infty$ we have ${\cal N} = \Phi,_r$,
and the components of $R_{ijkl}$ are finite as long as $|M,_r/\Phi,_r| <
\infty$.

\section{The general equation of an ERS}\label{genERS}

\setcounter{equation}{0}

We keep the assumptions $u = 0$ and $\varphi =$ constant along a null geodesic,
and take up the reasoning at (\ref{8.4}). Using (\ref{4.2}), Eq. (\ref{8.3}) can
be written as
\begin{eqnarray}\label{c.1}
&& \left(\frac {E,_r} {2 E} - \frac {{\cal E},_r} {\cal E}\right) \frac {\Phi^2
\Phi,_t} M \nonumber \\
&&\ \ \ - \left(\frac 3 2\ \frac {E,_r} E - \frac {M,_r} M\right) \left(t -
t_B\right) = - t_{B,r}.
\end{eqnarray}
Using (\ref{a.3}) in the above to eliminate $\Phi$, $\Phi,_t$ and
$M/(2E)^{3/2}$, one obtains
\begin{eqnarray}\label{c.2}
&& \left(t - t_B\right) \left[\left(\frac {E,_r} {2 E} - \frac {{\cal E},_r}
{\cal E}\right) F(\eta) + \frac {M,_r} M - \frac 3 2\ \frac {E,_r} E\right]
\nonumber \\
&& \ \ \ = - t_{B,r},
\end{eqnarray}
where
\begin{equation}\label{c.3}
F(\eta) \df \frac {(\cosh \eta - 1) \sinh \eta} {\sinh \eta - \eta}.
\end{equation}
The no-shell-crossing conditions imply $-t_{B,r} \geq 0$ \cite{HeKr2002}. We
have $F(\eta) > 3$ and $\dril F {\eta} > 0$ for all $\eta > 0$ and $\lim_{\eta
\to \infty} F(\eta) = \infty$. The coefficient of $F(\eta)$ in (\ref{c.2}) is
positive in consequence of (\ref{2.8}). The term $M,_r/M - (3/2) E,_r/E$ is not
guaranteed to be positive, but is independent of $\eta$, so, with sufficiently
large $\eta$ the first term in the square brackets will dominate over it and
will make the whole left-hand side of (\ref{c.2}) positive. Thus, with
sufficiently large $\eta$ (implying large $t - t_B$), the left-hand side of
(\ref{c.2}) will be larger than $(- t_{B,r})$, at any fixed $r$. On the other
hand, at $t = t_B$ (i.e. $\eta = 0$), the left-hand side of (\ref{c.2}) is zero,
so smaller than $(- t_{B,r})$. Consequently, at every $r$ (\ref{c.2}) has a
solution for $t$ somewhere in the range $t_B < t < \infty$, and so an ERS
exists.

\section{A Friedmann limit of the Szekeres spacetime}\label{frlim}

\setcounter{equation}{0}

Assume $M$ and $E$ to be of the Friedmannian forms (\ref{3.2}), (\ref{8.5}) and
$t_B =$ constant. Then (\ref{a.1}) -- (\ref{a.3}) imply
\begin{equation}\label{d.1}
\Phi(t, r) = r S(t);
\end{equation}
eq. (\ref{a.2}) shows this immediately, in (\ref{a.1}) and (\ref{a.3}) one first
has to note that $\eta$ is independent of $r$.

Next, substituting (\ref{3.2}) and (\ref{d.1}) in (\ref{2.5}) we obtain

\begin{equation}\label{d.2}
\kappa \rho = \frac {6M_0} {S^3}
\end{equation}
because the factor $1 - r {\cal E},_r/{\cal E}$ that appears in the numerator
and denominator of (\ref{2.5}) is nonzero and cancels out. Equation (\ref{d.2})
implies that $\rho$ depends only on $t$ in this limit. Thus, the set of
equations \{(\ref{3.2}), (\ref{8.5}), (\ref{d.1})\} is a sufficient condition
for the density to become spatially homogeneous, i.e. for (\ref{2.1}) to become
the Friedmann metric, with no conditions on $P$, $Q$ and $S$.

\section{Proof of (\ref{12.1}) with (\ref{12.3})}\label{condpq}

\setcounter{equation}{0}

With $P$ and $Q$ given by (\ref{12.2}), Eq. (\ref{12.1}) becomes
\begin{equation}\label{e.1}
\frac 1 {r^2} + \frac 1 {a^2 + r^2} - \frac {p^2 r^2} {\left(a^2 + r^2\right)^4}
- \frac {q^2 r^2} {\left(a^2 + r^2\right)^3} > 0.
\end{equation}
Denoting $r^2 = \chi$, this can be rewritten as
\begin{eqnarray}\label{e.2}
&&F(\chi) \df \left(a^2 + \chi\right)^4 + \chi \left(a^2 + \chi\right)^3 \nonumber \\
&&\ \ \ \ \ \ - q^2 \chi^2 \left(a^2 + \chi\right) - p^2 \chi^2 > 0.
\end{eqnarray}
This holds at all $\chi > 0$ if $p$ and $q$ obey (\ref{12.3}).

\section{Calculations underlying Figs. \ref{xyproj} -- \ref{yzproj}}\label{projcalc}

\setcounter{equation}{0}

\begin{figure}[h]
\includegraphics[scale=0.5]{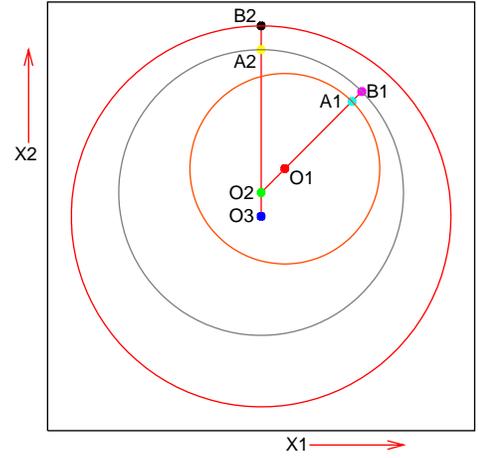}
\caption{Illustration to the calculations underlying Figs. \ref{xyproj} --
\ref{yzproj}. See text for explanations.} \label{circscheme}
\end{figure}

Figure \ref{circscheme} illustrates the numerical calculations underlying Figs.
\ref{xyproj} -- \ref{yzproj}. As with Fig. \ref{circles}, the mapping into the
Euclidean space was done so that the shortest distances between the spheres
(measured along the dipole maximum) were preserved. For clarity, the figure
shows the great circles of those spheres as if they all lied in the $(X_1, X_2)$
plane of the Euclidean space, but the calculations were done in 3 dimensions.

The 3 spheres (call them $S_1$, $S_2$ and $S_3$) have their centers at $O_1$,
$O_2$, $O_3$ and their radii are, respectively, the distances $\left|O_1
A_1\right|$, $\left|O_2 B_1\right|$ and $\left|O_3 B_2\right|$. The figure
illustrates how the calculation proceeds from sphere $S_1$ to $S_2$ and then
from $S_2$ to $S_3$.

The point $A_1$ is the position of the dipole maximum on sphere $S_1$. Given the
$r$-coordinate of $A_1$ (from the previous step of calculation), its $(x, y)$
coordinates are calculated from (\ref{12.5}) and then converted to the
$(\vartheta, \varphi)$ coordinates by (\ref{2.3}). The $(\vartheta, \varphi)$
coordinates are assumed to coincide with the spherical polar coordinates in the
Euclidean space of Figs. \ref{xyproj} -- \ref{yzproj}. (These coordinates are
defined even in the limit $r \to 0$ because (\ref{12.5}) has a well-defined
limit $r \to 0$.)

The radius $\left|O_1 A_1\right|$ is calculated using (\ref{9.5}). With $P(r)$
and $Q(r)$ being nonconstant, the path of ${\cal E},_r = 0$ is no longer
straight, but (\ref{9.7}) still holds, so (\ref{9.5}) still makes sense as the
definition of the geodesic radius. Then, $\left|A_1 B_1\right|$ is calculated
from (\ref{9.10}) using (\ref{12.6}) with $\varepsilon = +1$ for the value of
${\cal E},_r/{\cal E}$. At the point $B_1$, the $r$-coordinate of the sphere
$S_2$ is known, so we calculate its geodesic radius from (\ref{9.5}); this
radius is $\left|O_2 B_1\right|$.

Given the Euclidean $(X_1, X_2, X_3)$ coordinates of $O_1$ (also from earlier
calculation) and the distances $\left|O_1 A_1\right|$, $\left|A_1 B_1\right|$
and $\left|O_2 B_1\right|$ we calculate the position of the center of $O_2$ by
adding the quantity $\left(\left|O_1 A_1\right| + \left|A_1 B_1\right| -
\left|O_2 B_1\right|\right)$, projected on the $(X_1, X_2, X_3)$ Euclidean axes,
to the coordinates of $O_1$. Having thus determined the $(X_1, X_2, X_3)$
coordinates of $O_2$, we calculate the $(\vartheta, \varphi)$ coordinates of
$A_2$ (the dipole maximum on sphere $S_2$) from (\ref{12.5}) and (\ref{12.3}),
and proceed by the same method to determine the geodesic radius and the center
$O_3$ of the next sphere $S_3$.

The calculation begins by assuming a position for the sphere of radius $0$ (in
Figs. \ref{xyproj} -- \ref{yzproj} it is $X_1 = X_2 = X_3 = 0$). The centers and
radii of all the larger spheres are then calculated as described above.

This calculation involves an approximation, on top of the obvious approximation
connected with the numerical calculations. It assumes that the segments $A_1
B_1$ and $A_2 B_2$ are straight, so that ${\cal E},_r/{\cal E}$ along them
depends only on $r$ and is given by (\ref{12.6}).

\bigskip
\end{document}